\documentclass{article}
\usepackage{bm,latexsym,amsmath,amssymb,amsfonts,fancyhdr,color,graphicx,multirow,slashed,cite,bbold,cancel}
\usepackage[a4paper,bottom=3cm,top=2.5cm,head=0mm,width=17cm,dvipdfm]{geometry}
\usepackage[usenames,dvipsnames,svgnames,table]{xcolor}
\usepackage[colorlinks=true,
            linkcolor=blue,
            urlcolor=blue,
            citecolor=green,          
            bookmarks=true,
            bookmarksnumbered=true,
            breaklinks=true,
            pdfstartview=FitBH
            ]{hyperref}
\usepackage[dotinlabels]{titletoc}
\usepackage{titlesec}
\usepackage{authblk,ulem}

\numberwithin{equation}{section}
\allowdisplaybreaks[4]

\titlelabel{\thetitle.\quad \hspace{-0.8em}}
\titlecontents{section}
              [1.5em]
              {\vspace{4mm} \large \bf}
              {\contentslabel{1em}}
              {\hspace*{-1em}}
              {\titlerule*[.5pc]{.}\contentspage}
\titlecontents{subsection}
              [3.5em]
              {\vspace{2mm}}
              {\contentslabel{1.8em}}
              {\hspace*{.3em}}
              {\titlerule*[.5pc]{.}\contentspage}
\titlecontents{subsubsection}
              [5.5em]
              {\vspace{2mm}}
              {\contentslabel{2.5em}}
              {\hspace*{.3em}}
              {\titlerule*[.5pc]{.}\contentspage}

\newcommand{\titledef}{Neutrino Mass Measurement with Cosmic Gravitational Focusing} 

\hypersetup{ pdfauthor = {Shao-Feng Ge},
	     pdftitle = {\titledef}, 
	     pdfsubject = {}, 
             pdfkeywords = {}, 
	     pdfcreator = {LaTeX with hyperref package},
	     pdfproducer = {dvips + ps2pdf} }

\definecolor{gesfblack}{rgb}{0,0,0}

\definecolor{gesfblue}{rgb}{0.08,0.42,0.76}

\definecolor{gesfgreen}{rgb}{0,1,0}

\definecolor{gesfgrey}{rgb}{0.5,0.5,0.5}

\definecolor{gesflanse}{rgb}{0.00,0.50,0.50}

\definecolor{gesfpurple}{rgb}{0.47,0.19,0.42}

\definecolor{gesfred}{rgb}{1,0,0}

\definecolor{gesfwhite}{rgb}{1,1,1}

\definecolor{gesfyellow}{rgb}{0.7,0.4,0.3}

\newcommand{\gsec}[1]{{\hypersetup{linkcolor=red}Sec.\,\ref{#1}\hypersetup{linkcolor=blue}}}
\newcommand{\gapp}[1]{{\hypersetup{linkcolor=red}App.\,\ref{#1}\hypersetup{linkcolor=blue}}}
\newcommand{\geqn}[1]{\hypersetup{linkcolor=blue}Eq.\,(\ref{#1})\hypersetup{linkcolor=blue}}
\newcommand{\gfig}[1]{{\hypersetup{linkcolor=violet}Fig.\,\ref{#1}\hypersetup{linkcolor=blue}}}

\definecolor{Orange}{cmyk}{0,0.61,0.87,0}
\definecolor{JungleGreen}{cmyk}{0.99,0,0.52,0}
\definecolor{OliveGreen}{cmyk}{0.64,0,0.95,0.40}
\definecolor{Brown}{cmyk}{0,0.81,1,0.60}
\definecolor{RoyalBlue}{cmyk}{0.71,0.53,0,0.12}
\definecolor{Gray}{cmyk}{0,0,0,0.40}
\definecolor{LightPink}{cmyk}{0.0,0.25,0,0}
\definecolor{LLightPink}{cmyk}{0.0,0.10,0,0}
\definecolor{LightBlue}{cmyk}{0.25,0,0,0}
\definecolor{LightGray}{cmyk}{0,0,0,0.2}

\usepackage{float}

\graphicspath{{FinalFigures/}}

\begin{document}
\fontsize{12pt}{14pt}\selectfont

\title{
       \textbf{\fontsize{15pt}{16pt}\selectfont \titledef}} 
\author[1,2]{{\large Shao-Feng Ge} \footnote{\href{mailto:gesf@sjtu.edu.cn}{gesf@sjtu.edu.cn}}}
\affil[1]{Tsung-Dao Lee Institute \& School of Physics and Astronomy, Shanghai Jiao Tong University, Shanghai 200240, China}
\affil[2]{Key Laboratory for Particle Astrophysics and Cosmology (MOE) \& Shanghai Key Laboratory for Particle Physics and Cosmology, Shanghai Jiao Tong University, Shanghai 200240, China}
\author[3]{{\large Pedro Pasquini} \footnote{\href{mailto:pedrosimpas@g.ecc.u-tokyo.ac.jp}{pedrosimpas@g.ecc.u-tokyo.ac.jp}}}
\author[1,2]{{\large Liang Tan} \footnote{\href{mailto:tanliang@sjtu.edu.cn}{tanliang@sjtu.edu.cn}}}
\affil[3]{Department of Physics, University of Tokyo, Bunkyo-ku, Tokyo 113-0033, Japan}

\date{\today}

\maketitle

\begin{abstract}
\fontsize{12pt}{14pt}\selectfont

We thoroughly explore the cosmic gravitational focusing
of cosmic neutrino fluid (C$\nu$F) by dark matter (DM) halo
using both general relativity for a point
source of gravitational potential and Boltzmann
equations for continuous overdensities. Derived in
the general way for both relativistic and
non-relativistic neutrinos, our results
show that the effect has fourth power
dependence on the neutrino mass and temperature.
With nonlinear mass dependence which is different
from the cosmic microwave background (CMB) and
large scale structure (LSS) observations,
the cosmic gravitational focusing can provide an
independent cosmological way of measuring the neutrino
mass and ordering. We take DESI as an example to
illustrate that the projected sensitivity as well as its
synergy with existing terrestrial neutrino oscillation
experiments and other cosmological observations can
significantly improve the neutrino mass measurement.

\end{abstract}

\newpage

\section{Introduction}
\label{sec:intro}

When Pauli proposed neutrino to explain the continuous
$\beta$ decay spectrum in 1931, he also argued in the same
famous letter that neutrino mass should 
be no greater than 1\% of the proton mass
\cite{Pauli:1991owm}. Otherwise, the final-state electron
spectrum should have apparent distortion around its endpoint 
\cite{hornyak1950energy}. This feature can be used to
measure the absolute neutrino masses
\cite{hornyak1950energy,langer1952beta,Formaggio:2021nfz}
such as at the ongoing KATRIN experiment \cite{KATRIN:2001ttj}.
However, the $\beta$ decay endpoint measurement is sensitive
to the neutrino mass combination 
$m_\beta \equiv \sqrt{\sum_i |U_{ei}|^2 m_i^2}$
\cite{KATRIN:2021uub} where $U_{\alpha i}$ is the 
PMNS matrix element \cite{Workman:2022ynf}. While the
current limits are $m_\beta < 0.8$\,eV at
90\%\,C.L.\cite{KATRIN:2021uub},
the future sensitivity can reach $m_\beta \lesssim 0.2$\,eV
\cite{KATRIN:2005fny}.

Probably out of Pauli's expectation, not just the neutrino
itself but also its mass plays very
important roles in particle physics and cosmology.
As first established by atmospheric  
\cite{Super-Kamiokande:1998kpq} and solar
\cite{SNO:2002tuh} neutrino oscillation experiments,
neutrinos have non-zero mass squared differences 
$\Delta m_{ij}^2 \equiv m_i^2 - 
m_j^2$ \cite{Formaggio:2021nfz}.
The nonzero neutrino mass is the first experimentally
verified new physics beyond the Standard Model of particle
physics. Whether it is a genuine mass or just
environmental effect due to scalar \cite{Ge:2018uhz}
and dark \cite{Ge:2019tdi,Choi:2019zxy,Choi:2020ydp,Sen:2023uga} non-standard interactions
via forward scatterings needs further experimental
identification. However, the neutrino oscillation 
cannot measure the absolute neutrino mass. 
The current global fits
\cite{deSalas:2017kay,Capozzi:2018ubv,Esteban:2020cvm} 
of neutrino oscillation experimental data give
$\Delta m^2_s \equiv \Delta m_{21}^2 \approx 7.54 \times 10^{-5}$\,eV$^2$
and $|\Delta m^2_a \equiv \Delta m_{31}^2| \approx 2.47\times 
10^{-3}$\,eV$^2$\cite{Workman:2022ynf}. Note that the sign of $\Delta m^2_a$
is currently unknown which means the neutrino masses
can be either normal ordering (NO) if $\Delta m^2_a > 0$
with $m_1 < m_2 < m_3$
or inverted ordering (IO) if $\Delta m^2_a < 0$
with $m_3 < m_1 < m_2$. By 2030, the JUNO experiment
\cite{JUNO:2015zny,JUNO:2015sjr}
will determine the mass ordering
\cite{Petcov:2001sy,Choubey:2003qx,Learned:2006wy,Zhan:2008id,Zhan:2009rs,Qian:2012xh,Ge:2012wj,Qian:2012zn} with a
3$\sigma$ level of accuracy
and can further improve to
4$\sigma$ by also including the atmospheric neutrino
oscillation \cite{Zhang:2021adu}.

The neutrinoless double beta decay ($0 \nu 2 \beta$), 
$\mathcal N(A,Z)\rightarrow \mathcal N(A,Z + 2) + 2 e^-$
\cite{Furry:1939qr},
is another type of terrestrial experiment
to measure neutrino mass
if neutrinos are Majorana particles
\cite{Agostini:2022zub,Dolinski:2019nrj}.
Its half-life is  
proportional to the neutrino mass combination 
$m_{ee} \equiv \sum_i m_i U_{ei}^2$. 
Currently,
the best upper bounds come from Kamdland-ZEN 
\cite{KamLAND-Zen:2016pfg} and GERDA \cite{GERDA:2020xhi},
\begin{align}
  |m_{ee}|
\leq 
  6.1\times 10^{-2}~{\rm to}~1.65\times 10^{-1}\,{\rm eV}
\quad {\rm and} \quad 
  |m_{ee}|
\leq 
  7.9\times 10^{-2}~{\rm to}~1.8\times 10^{-1}\,{\rm eV},
\end{align}
at 90\% C.L. However, the effective mass $m_{ee}$ may
be less than $10^{-3} \,{\rm eV}$ in the NO which is
known as the funnel region. Although it seems
like the funnel region means null observation and
hence is experimentally unattractive, it allows pinning
down the absolution neutrino mass in the range of
$(2 \sim 6)$\,meV \cite{Ge:2015bfa}, simultaneously
measuring the two Majorana CP phases
\cite{Ge:2016tfx,Cao:2019hli}, and verifying whether
there is LMA-Dark solution \cite{Ge:2019ldu}.

For the cosmic neutrino background (C$\nu$B), its momentum
around only 1.9\,K is roughly the same size as the current
temperature of our Universe \cite{Dodelson:cosmo2nd}
and could be smaller than the neutrino masses. 
Consequently, the neutrino mass can
dominate the kinematics of C$\nu$B which allows
the possibility of distinguishing
the Dirac and Majorana nature by a factor of 2
\cite{Hernandez-Molinero:2022zoo}.
In addition, the PTOLEMY experiment
\cite{PTOLEMY:2018jst,PTOLEMY:2019hkd,Alvey:2021xmq}
aims at measuring the emitted electron when
capturing the cosmic relic neutrinos,
$\nu_e+{ }^3 \mathrm{H} \rightarrow e^{-} + { }^3 \mathrm{He}$.
While the peak heights are determined by the mixing
matrix elements, the peak locations of the electron
energy corresponds to the neutrino mass eigenvalues
\cite{Alvey:2021xmq}. If the electron energy resolution can
be significantly improved, it is possible to use C$\nu$B
for measuring the neutrino masses. Most notably, the
tiny mass can induce a factor of 2 difference for C$\nu$B
measurement via chiral oscillation
\cite{Bittencourt:2020xen,Ge:2020aen}.

Another phenomena that can measure neutrino mass is the
radiative emission of neutrino pair (RENP)
\cite{Yoshimura:2006nd,Fukumi:2012rn,Hiraki:2018jwu,Tashiro:2019ghs}.
With both superradiance and macroscopic
coherent material, the RENP process that is induced by weak
interactions can be significantly enhanced
\cite{Yoshimura:2011ri,Yoshimura:2012tm}.
Since the typical energy of atomic transitions is intrinsically
at the $\mathcal O($eV) scale, RENP can be sensitive to not just
light/massless mediators \cite{Ge:2021lur,Ge:2022cib,Ge:2023oag}
but also those properties that are attached to the
neutrino mass terms \cite{Yoshimura:2009wq,Dinh:2012qb}.
In particular, the $\mathcal (10^{-3} \sim 10^{-2})$\,eV
neutrino mass eigenvalues can leave a visible effect in the
associated photon spectrum and hence can be measured by
RENP \cite{Song:2015xaa,Zhang:2016lqp}. 

In addition to those terrestrial neutrino experiments
mentioned above, astrophysical observations can also use the
flight time delay of neutrinos from a nearby supernova to
measure the neutrino mass \cite{Zatsepin:1968kt}.
Massive neutrinos propagate slower than the speed of light
and hence experience flight time delay comparing with
the massless photon. This time delay depends on
the neutrino mass squared $m^2_\nu$,
$\Delta t = 0.005\,{\rm s} \times
(D/10\,{\rm kpc}) \times (m_\nu [{\rm eV}]/E_\nu 
[10{\rm MeV}])^2$ 
\cite{Zatsepin:1968kt,Hyper-Kamiokande:2018ofw}, 
where $D$ is distance and $E_\nu$ the neutrino energy. 
The observation of a few neutrino events
from SN1897A has led to the neutrino mass bound
$m_{\nu} < 5.7$\,eV 
at 95\% C.L. \cite{Loredo:2001rx}.
With the expectation of detecting several hundreds of
supernova neutrino events in the future, 
the neutrino mass limits can be pushed to
$m_\nu \lesssim 1$\,eV 
\cite{Lu:2014zma,Hyper-Kamiokande:2018ofw,Pompa:2022cxc}.

Cosmology is by far the most promising 
way to measure the neutrino masses
\cite{Lesgourgues:2006nd,Boyle:2017lzt}. 
Neutrinos are ultra-relativistic 
when decoupling from the thermal bath with 
temperature $\sim 1$\,MeV \cite{Lattanzi:2017ubx}. 
As the temperature drops, massive neutrinos become 
non-relativistic \cite{Lattanzi:2017ubx}. 
The non-relativistic neutrino decreases 
CMB power spectrum due to less ISW effect 
and suppress matter power spectrum 
due to the neutrino free-streaming
\cite{Workman:2022ynf}. 
Both CMB and LSS observations are sensitive to 
the neutrino energy density
$\rho_\nu \equiv \sum n_\nu m_i \approx n_\nu \sum m_i$
that is a product of the neutrino number density $n_\nu$
of a single species ($\nu_i$ and $\overline \nu_i$)
and the neutrino mass sum $\sum m_i$.
The combination of the Planck 2018 \cite{Planck:2018vyg}
and DES results \cite{DES:2021wwk} puts the most
stringent constraint, $\sum m_i < 0.13$\,eV at 95\% C.L.
\cite{Workman:2022ynf,DES:2021wwk}. Also, neutrino
clustering can bend the background photon to induce
a dipole structure in CMB 
\cite{Kaplinghat:2003bh,Hotinli:2023scz}. 
Similarly, the neutrino gravitational focusing
effect by DM halo can also be probed by the galaxy
weak lensing observation \cite{Zhu:2014qma}.
These two effects are sensitive to the neutrino masses.

In addition to the standard cosmology approaches,
we explore an additional and independent method for 
measuring neutrino masses based on the cosmic gravitational
focusing between the C$\nu$F
and dark matter halos \cite{Zhu:2013tma,Okoli:2016vmd,Zhu:2019kzb,Nascimento:2023ezc}.
With dependence on the mass fourth power instead
of the usual linear neutrino mass sum, the cosmic
gravitational focusing
is more sensitive to the neutrino masses and can
provide a complementary measurement. Such an effect
can be measured by the two-point and three-point galaxy correlation
observations at the coming next generation of galaxy
surveys. Especially, the Dark Energy Spectroscopic 
Instrument (DESI) \cite{DESI:2016fyo, DESI:2018ymu} has already 
started the data release \cite{DESI:2023ytc},
Euclid was launched on July 2023 \cite{EUCLID:2011zbd},
and CSST (China Space Station Telescope) will be
launched around 2024 or so 
\cite{cao2018testing,cao2022anisotropies,cao2022calibrating}. 
We will show the projected sensitivity by taking DESI
for illustration.

This paper is organized as follows. We introduce the
formalism for the focusing of C$\nu$F in the gravitational
potential of DM halo \gsec{sec:formalism} for both
non-relativistic and relativistic neutrinos.
In \gsec{sec:Observable}, we calculate the galaxy
correlation observables and the signal-to-noise
ratio (SNR). Finally in \gsec{sec:Numass},
we simulate the expected DESI sensitivity to the
neutrino mass with the galaxy categories summarized
in \gsec{sec:DESEresult} for illustration.
Our conclusion can be found in \gsec{sec:conclusion}.

\section{Cosmic Gravitational Focusing}
\label{sec:formalism}

When a flow of collisionless particles pass by a
massive object, they can experience gravitational
lensing \cite{DodelsonGL,CongdonKeeton,Meneghetti}.
With an attractive gravitational
force, particles with the same incoming velocity are
bent towards each other and focus behind the massive
object to enhance the density there. Such gravitational
focusing effect can leave detectable annual modulation in
the direct detection of DM with our Sun
being the massive object to provide gravitational source
\cite{Griest:1987vc,Sikivie:2002bj,Lee:2013wza,Bozorgnia:2014dqa}.
Since DM contributes the most to
the matter world in our Universe, cosmic gravitational
focusing of C$\nu$F under the influence of DM halo
happens more vastly. The C$\nu$F density increase
by the gravitational focusing will generate a dipole
density distribution
around the central DM halo which subsequently affects the
galaxy density distribution that can be observed
via the galaxy cross correlation.

We first derive the general expressions of cosmic
gravitational focusing with a single DM halo as a point
source of
gravitation force in \gsec{sec:force}. Especially,
we reproduce the dependence of drag force on the
fourth power of neutrino masses with non-relativistic
speed. However, our results apply more generally
for relativistic neutrino fluids with temperature
dependence. To overcome the intrinsic logarithmic
divergence in this point-source approach, we further
adapt the Boltzmann equation formalism in \gsec{sec:nugf}
which can apply more generally.

\subsection{The Trajectory and Drag Force Formalism}
\label{sec:force}

In addition to gravitational focusing, gravitational
lensing can also lead to dynamical friction
\cite{1943ApJ....97..255C} which plays important roles
in astrophysics, such as the formation of planets
\cite{2006Icar..184...39O}, star clusters
\cite{2003ApJ...597..312K}, and galaxies
\cite{1999PhR...321....1S}. So the cosmic gravitational
focusing of C$\nu$F under the influence of DM halo
is also called as dynamical friction \cite{Zhu:2013tma}.
Below we first provide a classical picture of single-particle
trajectory and its deflection to derive
the drag force which is an essential feature of
dynamical friction.

When an object moves through a bunch of collisionless
particles, its gravitational force can deflect the
originally free moving particles. To give an intuitive
picture, let us first consider a single neutrino passing 
by a DM halo with mass $M$. In the spherical coordinate
$(t,r,\theta,\phi)$
with DM halo at its origin, the space-time can be
described by,
\begin{eqnarray}
  ds^2
\equiv
- \left(1 - \frac{2GM}{r}\right)dt^2
+ \left(1 - \frac{2GM}{r}\right)^{-1} dr^2
+ r^2d\Omega^2,
\end{eqnarray}
where $G$ is the Newton constant and $d \Omega$ the
solid angle element. One may see that the effect of
the DM halo gravitational potential appears in front
of $d t^2$ and $d r^2$. A free moving neutrino follows
the geodesic equation that has four constants of motion,
\begin{subequations}
\begin{align}
&
  \left(1 - \frac{2GM}{r}\right)\frac{dt}{d\tau}
\equiv
  C_t,
\quad 
  r^2 \frac{d\theta}{d\tau}
\equiv
  C_\theta,
\quad 
  r^2\sin^2 \theta \frac{d\phi}{d\tau}
\equiv
  C_\phi,
\\
&
  \left(1 - \frac{2 G M} r \right)^{-1}
  \left(\frac{d r}{d \tau}\right)^2 
-
  \left(1 - \frac{2 G M} r \right)^{-1}
  C_t^2
+
  \frac{C_\phi^2}{r^2}
\equiv
  C_r,
\end{align}
\label{eq:geodesic}
\end{subequations}
where $\tau$ is the proper time. The integration
constants $C_t$, $C_r$, $C_\theta$, and $C_\phi$
are determined by initial conditions. Due to reflection
symmetry, one may reduce the three-dimensional
coordinate to a two-dimensional one where the neutrino
trajectory resides on the $x-y$ plane with $z$
always being 0. This means the zenith angle is
fixed with $\theta = \pi/2$ and $C_\theta = 0$. In addition
to time $t$, both the radius $r$ and azimuthal angle $\phi$
can vary. Taking the radius $r$ to infinity, $r\rightarrow \infty$,
the metric approaches a free particle and we can identify 
$C_t = \sqrt{m_\nu^2 + |\bm p_i|^2}/m_\nu \equiv E_i/m_\nu$, 
$C_\phi = E_i |\bm b| |\bm v_i|/m_\nu$, 
and $C_r = -1$, 
where $\bm b$ is the impact parameter and 
$\bm v_i \equiv \bm p_i/E_i$ is the neutrino initial velocity.

Combining the equations for $\phi$ and $r$ in \geqn{eq:geodesic}
with the relation $d\phi/dr = d\phi/d\tau (dr/d\tau)^{-1}$,
we obtain the total deflection angle of the neutrino 
when passing by the DM halo,
\begin{eqnarray}
  \Delta \phi 
=
  2|{\bm b}|
  \int_{r_{\rm min}}^{\infty}
  \frac{dr}{r^2}
\left[
    1
    + \frac{2 b_{90}}{r}
    - \frac{|{\bm b}|^2}{r^2}
    +\frac{2GM|{\bm b}|^2}{r^3}
\right]^{-1/2}.
\label{eq:int_dphi}
\end{eqnarray}
Note that $b_{90} \equiv G M m_\nu^2 /|\bm p_i|^2$
denotes the impact parameter that corresponds to
a $90^\circ$ deflection of the neutrino momentum.
Given impact parameter $|{\bm b}|$, the closest distance
between neutrino and the central DM halo is $r_{\rm min}$
as solution of 
$r^3_{\rm min}
+2 b_{90} r^2_{\rm min} -|{\bm b}|^2 r_{\rm min}
+2 GM |{\bm b}|^2 = 0$.
For the large impact parameter approximation,
$|\bm b| \gg b_{90}$ and $GM$,
the total deflection angle between the initial and
final neutrino position vectors can be perturbatively
expanded as,
\begin{align}
  \Delta \phi 
\approx
  \pi 
+ 2 \frac{GM}{|{\bm b}|}
    \left(\frac{m_\nu^2}{|\bm p_i|^2} + 2 \right).
\label{eq:Dphi}
\end{align}
The deflection angle of the neutrino momentum corresponds
to $\Delta \phi - \pi$.
We take the $x$-direction anti-parallel to the initial
neutrino velocity in the DM frame. Then, the initial 
$\hat {\bm p}_i$ has $\phi_i= \pi$
with direction $\hat {\bm p}_i = 
- \hat {\bm x}$ while the final momentum
$\bm p_f \equiv |{\bm p}_i| 
(\cos \phi_f \hat {\bm x} + \sin \phi_f {\hat {\bm y}})$
has $\phi_f = \Delta \phi$.

We consider a stream of neutrinos whose phase space
distribution $f({\bm p})$ is spatially independent.
In other words, for any neutrino with certain momentum
and impact parameter ${\bm b}$ to pass by a DM halo,
there is equal chance for a neutrino with the same
momentum but opposite impact parameter $-{\bm b}$.
Such a pair of neutrinos would experience equal amount
of bending by the DM halo gravitational potential
in opposite directions. In total, their total momentum along the
initial one ($\hat{\bm p}_i = - \hat{\bm x}$) is reduced
while no total perpendicular (transverse) component can develop.
The average momentum  transfer is parallel to the $x$-direction 
and can be obtained from \geqn{eq:Dphi} as, 
\begin{align}
  \Delta{\bm p}^{\parallel} 
\equiv  
  \bm p_f^{\parallel} - \bm p_i 
=
  \left(- \cos \Delta \phi -1 \right)
  \bm p_i
\approx   
- \frac{2 G^2 M^2}{ |\bm b|^2}
  \left(
    \frac{m_\nu^2}{|\bm p_i|^2} + 2 
  \right)^2 
  \bm p_i.    
\end{align}

To calculate the drag force, let us consider a bulk of
neutrino with roughly the same initial velocity ${\bm v}_i$
(equivalently momentum ${\bf p}_i$) and magnitude of
impact parameter $|{\bm b}|$,
\begin{eqnarray}
  d N
=
  (2\pi |\bm b| d|\bm b|)
  (|{\bm v}_i| \Delta t)
  d n_{{\bm p}_i}.
\label{eq:dN}
\end{eqnarray}
The first parenthesis $2\pi |\bm b| d|\bm b|$ gives the
cross section around the impact parameter $|{\bf b}|$ while
the second $|{\bm v}_i| \Delta t$ is the distance traveled
by neutrinos within the time during $\Delta t$. Together,
the product of volume
$dV \equiv (2\pi |\bm b| d|\bm b|) (|{\bm v}_i| \Delta t)$
and the particle number density
$d n_{{\bm p}_i} \equiv f_\nu({\bm p}_i)d^3{\bm p}_i/(2 \pi)^3$
in the momentum volumn $d^3{\bm p}_i$ gives the particle number $d N$.

The drag force on DM halo is then an integration
of $\Delta {\bm p}^\parallel / \Delta t$ over $d|\bm b|$
and $d^3{\bm p}_i \equiv |{\bm p}_i|^2 d |{\bm p}_i| d \Omega_{{\bm p}_i}$,
\begin{align}
  \bm F_{\rm drag}
\equiv 
  \int
  \frac{\Delta {\bm p}^\parallel}{\Delta t}
  d N 
=
  \frac{2}{\pi}
  \int
  \frac{G^2 M^2 d|\bm b|}{|\bm b|}
  \int 
  \frac{d|\bm p_i|}{E_{{\bm p}_i}}
  \left(m_\nu^2+2|\bm p_i|^2\right)^2
  \int \frac{d\Omega_{\bm p_i}}{4\pi}
  f_\nu(\bm p_i)\hat{\bm p_i},
\label{eq:FfricInt}
\end{align}
where the neutrino velocity $|{\bm v}_i|$ in \geqn{eq:dN}
has been replaced by $|{\bm p}_i| / E_{{\bf p}_i}$.
For clarity, the integration has been divided into
three parts for the impact parameter $d|{\bm b}|$,
the neutrino momentum $d |{\bm p}_i|$, and the neutrino
direction $d \Omega_{{\bm p}_i}$. If there is no
relative velocity between the C$\nu$F and DM halo,
the neutrino phase space distribution is isotopic,
$f_\nu({\bm p}_i) = f_\nu(|{\bm p}_i|)$, and consequently
there is no dynamical friction. In other words,
the dynamical friction comes from a nonzero relative
velocity ${\bm v}_{\nu c}$ or equivalently an
anisotropic phase space distribution.

Starting with the Fermi-Dirac phase space distribution
in the C$\nu$F frame,
$f_\nu(\bm p') =  2 / ( e^{|\bm p'|/T} + 1 )$
including both the neutrino and antineutrino contributions,
we perform a Lorentz boost to the DM halo frame. 
Since the relative velocity between the two frames is small 
$|\bm v_{\nu c}| \sim \mathcal O(10^{-3})$ \cite{Zhu:2013tma}
as elaborated in \gapp{sec:Vnuc}, 
the Lorentz boost can be expanded to the linear order of
$\bm v_{\nu c}$,
$\bm p' \approx  \bm p_i - E_{\bm p_i} \bm v_{\nu c}$.
Consequently, the neutrino phase space distribution becomes,
\begin{eqnarray}
  f_\nu(\bm p_i, \bm v)
\approx
  \frac 2 {e^{ | \bm p_i - E_{\bm p_i} \bm v_{\nu c} | / T} + 1 }.
\label{eq:nuPhase-Space}
\end{eqnarray}
Together with \geqn{eq:FfricInt}, the drag force is
also linearly proportional to $\bm v_{\nu c}$,
\begin{align}
  \bm F_{\rm drag}
=
  \frac{2 {\bm v}_{\nu c}G^2}{3\pi}
\left(
  m^4_\nu    
+ \frac{4\pi^2}{3}m_\nu^2 T_\nu^2 
+ \frac{28 \pi^4}{15} T_\nu^4
\right)
  \int M^2(<|{\bm b}|) \frac{d|{\bm b}|}{|{\bm b}|}. 
\label{eq:Force_drag}
\end{align}
For non-relativistic neutrinos, $m_\nu \gg T_\nu$,
the drag force is proportional to the fourth power
of the neutrino mass $m_\nu^4$ which is a unique feature
of the effect \cite{Okoli:2016vmd}.
For the relativistic case, $m_\nu \ll T_\nu$, the drag force 
is proportional to the fourth power of its temperature $T_\nu^4$.
This means that even massless neutrinos can induce a
nonzero drag force on a DM halo.

However, the classical trajectory and force description
summarized here has intrinsic flaws.
When the impact parameter $\bm b$ goes to infinity or zero,
the integration receives logarithmic divergence \cite{Okoli:2016vmd}.
Although the zero limit can be naturally avoided since
the DM halo mass distribution $M(< |{\bm b}|)$ is not a
point source but instead has extended geometry, the other limit
$|{\bm b}| \rightarrow \infty$ has intrinsic difficulty
with a stationary gravitational potential which effectively
treat the gravitational attraction as force at a distance.
This is not true since the DM halo forms and develops
together with the evolution of our Universe.
In addition, the
key factor that affects the observable galaxy
cross correlation function as elaborated in
the next section is actually the focused density rather
than the drag force. So it is necessary to find a more
appropriate description of the cosmic gravitational
focusing effect.

\subsection{The Boltzmann Equation Formalism}
\label{sec:nugf}

The evolution of DM halo and the large scale structure
of the whole Universe can be naturally described by
the Boltzmann equations. Although there is no geometric
picture as clear as the trajectory and drag force formalism
in the previous \gsec{sec:force}, the Boltzmann equation
formalism better fits the continuous overdensity
distribution and cosmic gravitational focusing.
Intuitively thinking,
when the C$\nu$F passes by a DM halo, 
the bending of their trajectories also changes
the neutrino density distribution. In other words,
the neutrino density distribution can also manifest
the trajectory bending and drag force.

The evolution of the neutrino phase space distribution
is described by the following Boltzmann equation,
\begin{align}
\Biggl\{
  \partial_t 
+
  \frac {\bm p \cdot {\nabla}_{\bm x}} {a E_{\bm p}}  
-
\left[ 
  (H + \dot{\Phi}) {\bm p}
+ \frac{E_{\bm p}}{a} \nabla_{\bm x}\Psi
- \frac {|\bm p|^2 \nabla_{\bm x} \Phi
         - {\bm p} (\bm p \cdot \nabla_{\bm x} \Phi)}
          {a E_{\bm p}}
  \right]
  \cdot
  \nabla_{\bm p}
  \Biggr\} f_\nu (\bm x, \bm p)
= 0,
\label{eq:Boltz_eq}
\end{align}
where the gravitational effect of the DM halo has been
encapsuled in the gravitational potentials $\Phi$ and $\Psi$
as defined in the conformal-Newtonian gauge 
$ds^2 = - (1 + 2 \Psi) dt^2 + (1 + 2 \Phi) a^2 d{\bm x}^2$
with $a$ being the scale factor
\cite{Dodelson:cosmo2nd}.
The second term originally has a coefficient
$1 - \Phi + \Psi$ that is approximately 1 since both
$\Phi$ and $\Psi$ are negligibly small. For further
simplicity, one may also neglect the anisotropic stress,
$\Phi = -\Psi$.
The neutrino energy $E_{\bm p}$ is subject to the  
on-shell condition, 
$E_{\bm p} = \sqrt{|\bm p|^2 + m_{\nu}^2}$.

We consider the late stage of Universe evolution
when the cosmic gravitational focusing of C$\nu$F under
the influence of DM halo becomes important. 
For simplicity, the C$\nu$F has static phase space distribution,
$\partial_t f_\nu \approx 0$, and the Hubble rate $H$ does
not play a major role at small scales. Comparing the
Hubble parameter term and the second term in the same
bracket of \geqn{eq:Boltz_eq}, the above approximation
happens at 
$|\bm p| / m_\nu \gg H \Delta x$, namely,
$\Delta x \ll 7.4\,{\rm Mpc}\,(0.1 {\rm eV}/m_\nu) (1+z)^{-1/2}$ 
\cite{Okoli:2016vmd} which is exactly 
the scale for neutrino free streaming as explored in
\cite{Nascimento:2023ezc}.
In addition, the DM halo gravitational
potential also varies slowly with time, $\dot \Phi \sim 0$.
The neutrino phase space distribution, $f_{\nu}({\bm x}, {\bm p})$, 
is then described by the time-independent Boltzmann equation
\cite{Dodelson:cosmo2nd},
\begin{eqnarray}
\Biggl\{
  \frac{1}{aE_{\bm p}}  
  {\bm p}\cdot {\nabla}_{\bm x}
-\left[ 
    \frac{E_{\bm p}}{a}
    \nabla_{\bm x}\Psi
+ \frac {|\bm p|^2 \nabla_{\bm x} \Psi
         - {\bm p} (\bm p \cdot \nabla_{\bm x} \Psi)}
          {a E_{\bm p}}
  \right]
  \cdot\nabla_{\bm p}
\Biggr\}
  f_{\nu}({\bm x}, {\bm p})  
= 0.
\label{eq:General_Boltzmann_eq}
\end{eqnarray}
In \cite{Nascimento:2023ezc},
a more general derivation beyond the static Boltzmann approximation is provided.

To make the focusing effect more apparent, the neutrino
phase-space distribution can be decomposed into 
an homogeneous part $\overline f_\nu(\bm p)$ plus a tiny deviation 
$\delta f_\nu$,
\begin{align}
  f_\nu(\bm x, \bm p) 
\equiv 
  \overline f_\nu(\bm p) 
+ \delta f_\nu(\bm x, \bm p).
\end{align}
Putting the decomposition back into the Boltzmann equation 
\geqn{eq:General_Boltzmann_eq}, the deviation $\delta f_\nu$
can be induced by its homogeneous counterpart $\overline f_\nu$
in the Fourier space,
\begin{eqnarray}
\delta \widetilde f_{\nu}({\bm k}, {\bm p})
=
  \widetilde \Psi({\bm k})
\left(
  \frac{m_\nu^2 + 2{\bm p}^2}
    {\bm p \cdot \bm k}
  \bm k
- \bm p
\right)
  \cdot 
  \nabla_{\bm p}
  \overline f_{\nu}({\bm p}).
\label{eq:deltaf_f}
\end{eqnarray}
The quantities with a tilde are the Fourier modes,
$\delta \widetilde f_\nu (\bm k, \bm p)
\equiv 
  \int d^3 \bm x 
  e^{ - i \bm k \cdot \bm x}
  \delta f_\nu(\bm x, \bm p)$.
We can see that the gravitational potential $\widetilde \Psi$
induces the inhomogeneous deviation $\delta \widetilde f_\nu$
which is proportional to $\widetilde \Psi$ at the leading order.

The neutrino energy density fluctuations are obtained 
by integrating $\delta \widetilde f$ over all possible
neutrino momenta,
\begin{align}
  \delta \widetilde \rho_\nu (\bm k)
\equiv 
  \int \frac{d^3 {\bm p}}{(2\pi)^3} E_{{\bm p}} 
  \delta \widetilde f_{\nu}({\bm k}, {\bm p})
=
  \widetilde \Psi({\bm k})
  \int \frac{d^3 {\bm p}}{(2\pi)^3} 
  E_{\bm p} 
\left(
  \frac{m_\nu^2 + 2{\bm p}^2}
    {\bm p \cdot \bm k}
  \bm k
- \bm p
\right)
  \cdot 
  \nabla_{\bm p}
  \overline f_{\nu}({\bm p}).
\label{eq:drho_nu}
\end{align}
If there is no relative velocity, the neutrino phase space distribution is isotropic,
i.e. $\overline f_\nu (\bm p)  = \overline f_\nu (|\bm p|)$. 
Then, the momentum gradient is proportional to the neutrino
momentum, $\nabla_{\bm p} \overline f_{\nu}(|{\bm p}|)
\propto {\bm p}$. Consequently, the wave number ($\bm k$)
dependence in the parenthesis disappears,
$\delta \widetilde \rho_\nu (\bm k) \propto \widetilde
\Psi({\bm k}) \int E^3_{\bm p} d^3 {\bm p}$, and the only
$\bm k$ dependence follows the DM gravitational potential
$\widetilde \Psi({\bm k})$. In other words, the neutrino
density fluctuation $\delta \widetilde \rho_\nu$ should
following the DM one if there is no relative velocity
and there is no dipole density distribution on the upstream
and downstream sides around the DM halo. In the presence
of relative velocity,
the cosmic gravitational focusing should lead to larger neutrino
density on the downstream side.

\subsubsection{Neutrino Density Anisotropy from the Relative Velocity}
\label{sec:trajectory}

In the presence of relative velocity, the neutrinos
deflected by a DM halo tend to accumulate downstream.
As a result, the neutrino density fluctuation has an
anisotropy, $\delta \rho_\nu(\bm x) \neq \delta \rho_\nu(-\bm x)$
along the relative velocity. Correspondingly, the
neutrino density fluctuation in the wave number space
after Fourier transformation would possess a non-zero
imaginary part, i.e.
${\rm Im} [\delta \widetilde \rho_\nu (\bm k)] \neq 0$
\footnote{
For any function with dipole structure, $A(-\bm x) = - A(\bm x)$,
its Fourier transformation 
$\widetilde A (\bm k) \equiv \int d \bm x e^{- i \bm k \cdot \bm x} A(\bm x)$
is a purely imaginary number
since
$[\widetilde A (\bm k) ]^* 
= 
\int d \bm x e^{ i \bm k \cdot \bm x} A(\bm x)
= 
\int d \bm x e^{- i \bm k \cdot \bm x} A(- \bm x)
=
-
\int d \bm x e^{- i \bm k \cdot \bm x} A( \bm x)
=
- \widetilde A (\bm k)
$.
}.

To find the explicit form of ${\rm Im} [\delta \widetilde \rho_\nu]$, 
we use the anisotropic phase space distribution
in \geqn{eq:nuPhase-Space}
as $\overline f_\nu = f(\bm p, \bm v_{\nu c})$ back
into \geqn{eq:drho_nu}. With a nonrelativistic
${\bm v}_{\nu c}$, it is much more convenient to
first expand the distribution function,
$f_\nu(\bm p, \bm v_{\nu c}) 
\approx
  2 /(e^{|\bm p|/T_\nu} + 1)
+
  E_{\bm p} \hat {\bm p} \cdot \bm v_{\nu c}
  /T_\nu (1 + \cosh |\bm p|/T_\nu)
$.
Consequently, the momentum gradient becomes, 
$ \nabla_{\bm p} \overline f_\nu 
\approx 
  \hat{\bm p}' [1-( \bm v_{\nu c} \cdot \hat{\bm p}' )/{E_{\bm p}'}]
  d \overline f (|\bm p'|)$ $/ d|\bm p'|$,
by noticing
that the neutrino distribution 
$\overline f_\nu (|\bm p'|)$ 
depends only on the modulus of the variable 
$\bm p' = \bm p - E_{\bm p_i} \bm v_{\nu c}$.
Then, \geqn{eq:drho_nu} becomes,
\begin{align}
\hspace{-0.6cm}
  \delta \widetilde \rho_\nu (\bm k)
=
  \widetilde \Psi({\bm k})
  \int \frac{d^3 {\bm p}}{(2\pi)^3} E_{\bm p} 
\left(
  \frac{m_\nu^2 + 2{\bm p}^2}{ \bm p \cdot \bm k }
  {\bm k} \cdot \hat{\bm p}'
- {\bm p} \cdot \hat{\bm p}'
\right)
\left(
  1
- \frac{{\bm v}_{\nu c} \cdot \hat {\bm p}'}{{E_{\bm p}'}}
\right)
  \frac{d\overline f_{\nu}(|\bm p'|)}{d|{\bm p}'|}.
\label{eq:drho_nu2}
\end{align}
Now, we shift the variable of integration by 
$\bm p \rightarrow \bm p' + E_{\bm p} \bm v_{\nu c}$ 
to write \geqn{eq:drho_nu2} as,
\begin{align}
\hspace{-3mm}
  \delta \widetilde \rho_\nu (\bm k)
\approx 
  \widetilde \Psi
  \int \frac{d^3 {\bm p}' }{(2 \pi)^3}
  \frac{d \overline f_\nu}{d |{\bm p}'|} 
  \frac{
   E_{{\bm p'}}^3}{( {\bm p}' + E_{{\bm p'}} {\bm v}_{\nu c} ) \cdot {\bm k}}
\left\{
    {\bm k} \cdot \hat {\bm p}' 
  - \frac{|{\bm p}'|}{E_{{\bm p}'}^2}
    \left[  
       {\bm v}_{\nu c} \cdot {\bm k}
     - 4 ( \hat {\bm p}' \cdot {\bm v}_{\nu c})
      ({\bm k} \cdot \hat {\bm p}') 
  \right]
\right\},
\label{eq:RelrhoInt}
\end{align}
where we ignore terms of order $\mathcal O (|\bm v_{\nu c}|^2)$.
While $\bm p$ is defined in the DM halo frame, its
counterpart $\bm p'$ is in the C$\nu$F frame.

Notice that the denominator 
$( \bm p' + E_{\bm p}' \bm v_{\nu c} ) \cdot \bm k$ has a pole 
when $\bm p' \cdot \bm k = - E_{\bm p'} \bm v_{\nu c}\cdot \bm k$. 
It is this pole that generates the imaginary 
part of $\delta \widetilde \rho_\nu$. With overdensity, the
gravitational potential always has a limited size and cannot
extend to infinity. Correspondingly, an imaginary part of the
wavenumber, $\bm k \rightarrow \bm k  -  i \epsilon$, 
is introduced in the denominator for
regularization \cite{Okoli:2016vmd}.
Then, the Sokhotski-Plemelj-Weierstrass (SPW) theorem \cite{Weinberg:1995mt}
can be used to extract the imaginary part, 
\begin{eqnarray}
{\rm SPW ~ theorem} \quad : \quad 
  \lim_{\epsilon \rightarrow 0^+}
  \frac{1}{x - i \epsilon }
=
  \mathcal P\left(\frac 1 x\right) 
+ i \pi \delta(x),
\label{eq:SPW_teo}
\end{eqnarray}
with $\mathcal P(1/x)$ indicating the principal value of the function $1/x$.
It is clear that the imaginary part arises from the $\delta$-function 
and is non-zero if the angle between ${\bm p}'$ and $\bm k$ 
is $\cos \theta \equiv \hat {\bm p}' \cdot \hat {\bm k} = - E_{\bm p'}\bm v_{\nu 
c} \cdot \hat {\bm k} / \bm p'$ for $|{\bm p'}| \geq 
E_{{\bm p'}} |{\bm v}_{\nu c} \cdot \hat {\bm k}|$.
From \geqn{eq:RelrhoInt},
we obtain the imaginary part of the neutrino density fluctuations, 
\begin{align}
  {\rm Im}[\delta \widetilde \rho_\nu]
= &
-
 \frac{ ({\bm v}_{\nu c} \cdot \hat {\bm k})\widetilde \Psi}{4\pi}
  \int d|{\bm p'}| 
\left(
    m_\nu^4
  + 3 m_\nu^2|{\bm p}'|^2
  + 2 |{\bm p}'|^4
\right)
  \frac{d \overline f_\nu}{d |{\bm p}'|}
    \Theta(|{\bm p}'| - E_{{\bm p'}} |{\bm v}_{\nu c} \cdot \hat {\bm k}|),
\label{eq:ImrhoPsi} 
\end{align}
The derivation details can be found in \gapp{app:phase_form}.
Here, we use the on-shell condition $E_{\bm p'}^2 = |\bm p'|^2 + m_\nu^2$,
and the Heaviside-$\Theta$ function to ensure that
$|{\bm p}'| \geq E_{{\bm p'}} |{\bm v}_{\nu c} \cdot \hat {\bm k}|$.
As we shall see in the next section, 
only the imaginary part of density fluctuations
is relevant for the dipole galaxy correlation.
Thus, calculating the corresponding real part is unnecessary.
\geqn{eq:ImrhoPsi} explicitly 
shows that the non-zero relative velocity $\bm v_{\nu c}$ 
leads to an imaginary part of neutrino energy density 
fluctuations in the Fourier space $\delta \widetilde \rho_\nu$.

\subsubsection{Matter Density Anisotropy Contributed by C$\nu$F}
\label{sec:deltam}

The fluctuations in the neutrino density 
contribute to the total matter density fluctuation 
that directly affects galaxy formation. Although cosmic
neutrinos are usually treated as radiation, they can be
both relativistic or non-relativistic in the late Universe
depending on the mass eigenvalues. For heavy neutrinos
that are already non-relativistic around redshift $z = 10$,
they can be readily counted as matter. Although light
neutrinos are still relativistic and hence should belong
to radiation, its energy density is much smaller than the
heavy neutrinos and hence it does not hurt to also
count them as matter. For convenience, we define the
sum of matter and neutrino densities as the total matter
density and denote the corresponding overdensity as
$\widetilde \delta_m$ in contrast to the genuine matter
one $\widetilde \delta_{m0}$.

Since the cosmic gravitational focusing leaves an imaginary
correction in the neutrino density fluctuation,
the total matter overdensity,
$\widetilde \delta_m \equiv \delta \widetilde \rho_m
/\rho_m$, is shifted from its real value, 
$\widetilde \delta_{m0}$($ \equiv \delta \rho_{m_0}/\rho_m$), 
to a complex one, 
$\widetilde \delta_{m0} \rightarrow 
\widetilde \delta_m \equiv \widetilde \delta_{m0}(1 + i \widetilde \phi)$. 
Here we have used $\widetilde \phi$ to parametrize the
relative imaginary part while the magnitude is combined
into $\widetilde \delta_{m0}$.
More concretely, the total matter overdensity is a sum of
the matter overdensity 
$\delta_{m0}$ and neutrino overdensities $\delta_{\nu_i}$,
\begin{align}
 \widetilde \delta_m 
\equiv 
  \widetilde \delta_{m 0}
  (1 + i \widetilde \phi)
=
  \widetilde \delta_{m0}
+ \sum_{i=1}^3 \frac{\delta \widetilde \rho_{\nu_i}}{\rho_m}
\equiv
  F_c \widetilde \delta_c
+ F_\nu \widetilde \delta_\nu,
\label{eq:dmgeneral}
\end{align}
where the first term is contributed mainly by DM
while the second by C$\nu$F. Note that
$\widetilde \delta_m$ is not simply a sum of
the DM relative overdensity $\widetilde \delta_c$
and the neutrino one $\widetilde \delta_\nu$.
The weighting factors $F_c$ and $F_\nu$
\cite{LoVerde:2014pxa} accounts for the fact
that the total overdensity is defined as,
$\delta_m \equiv (\delta \rho_c + \delta \rho_\nu)
/ (\rho_c + \rho_\nu)
\equiv 
  F_c \delta_c + F_\nu \delta_\nu$
with $F_c \equiv \rho_c / (\rho_c + \rho_\nu)$
and $F_\nu \equiv \rho_\nu / (\rho_c + \rho_\nu)$.
 
To obtain a neat form of $\widetilde \phi$,
we put \geqn{eq:ImrhoPsi} into \geqn{eq:dmgeneral}
and replace the gravitational potential with
the matter overdensity according to the Possion equation, 
$\widetilde \Psi 
= - 4 \pi G a^2 \rho_m \widetilde \delta_{m0}/|{\bm k}|^2$ 
\cite{Dodelson:cosmo2nd}. 
The matter density $\rho_m$ in $\widetilde \Psi$ cancels with the 
denominator in \geqn{eq:dmgeneral} and
the overall factor $\delta_{m0}$ is also factorized out,
\begin{align}
  \widetilde \phi 
=
  \frac{G a^2 }{|\bm k|^2}
  \sum_i ({\bm v}_{\nu_i c} \cdot \hat {\bm k})
  \int d|{\bm p'}| 
\left(
    m_i^4
  + 3m_i^2|{\bm p'}|^2
  + 2|{\bm p'}|^4
\right)
  \frac{d \overline f_\nu}{d |{\bm p'}|}
    \Theta(|{\bm p'}| - E_{{\bm p'}} |{\bm v}_{\nu_i c} \cdot \hat {\bm k}|),
\label{eq:phiGeneral}   
\end{align}
where the phase space distribution 
$\overline f_\nu(|\bm p'|) = 2 / [ e^{|\bm p'|/ T} + 1 ] $ 
includes both the neutrino and anti-neutrino contributions.
The $1/|\bm k|^2$ dependence holds a clear physical significance.  
Below the neutrino free-streaming scale $k_{\rm fs}$,
the neutrino overdensity is suppressed by a factor $( k_{\rm fs} /|\bm k|)^2$
compared to the matter overdensity $\delta_m$. 
To extend beyond the static Boltzmann approximation, 
we can simply replace the factor $ (k_{\rm fs} / |\bm k|)^2  $
with $ (k_{\rm fs} / |\bm k|)^2 / ( 1 + k_{\rm fs} / |\bm k|)^2 $
\cite{Nascimento:2023ezc}.
This provides a more general formalism than our
approach
by incorporating a $\Theta$ function in \geqn{eq:vnuc_Theta},
with roughly the same result.

Since the square root of two neutrino mass differences, 
$\sqrt{\Delta m_{21}^2}\approx 8.7\times10^{-3}$\,eV 
and $\sqrt{\Delta m_{31}^2}\approx 5\times10^{-2}$\,eV
measured by neutrino oscillation experiments \cite{Workman:2022ynf}, 
are much larger than the current neutrino 
temperature, $T_\nu \approx 1.7 \times 10^{-4}$\,eV,
at least one neutrino is non-relativistic for $z \lesssim 10$.
In this approximation, $|\bm p| \ll m_\nu$, the second and third terms in the parenthesis of  
\geqn{eq:phiGeneral} can be ignored 
to give \cite{Okoli:2016vmd},
\begin{align}
   \widetilde \phi 
\approx &
- \frac{2 G a^2}{|\bm k|^2}
  \sum_i 
  \frac{m_i^4 ({\bm v}_{\nu_i c} \cdot \hat {\bm k})}{e^{m_i |\bm v_{\nu_i c}\cdot \hat{\bm k}|/T_\nu} + 1}.
\label{eq:phiNonRel}  
\end{align}
To the leading order, $\widetilde \phi$ is proportional to 
the fourth power of neutrino mass, $\sum m_i^4$,
which is the same as the drag force in \geqn{eq:Force_drag}.
This is a prominent feature of the cosmic gravitational focusing 
and contrasts with the usual CMB and LSS constraints that
are sensitive to the neutrino mass sum, $\sum m_i$ \cite{Workman:2022ynf}. 
Later, we show that their combination improves the
sensitivity of cosmological constraints on the neutrino mass
in \gsec{sec:Numass}.

Our result in \geqn{eq:phiGeneral} describes both relativistic and non-relativistic neutrinos. For relativistic neutrinos, we write \geqn{eq:phiGeneral} 
in terms of special functions with a variable change  
$y \equiv | {\bm p'}|/T_\nu$,
\begin{align}
  \widetilde \phi
= &
  \frac{G a^2}{|\bm k|^2}
  \sum_i
  ({\bm v}_{\nu_i c} \cdot \hat {\bm k})
\left[ 
      m_i^4 f_0(y_i)
    + 3 m_i^2 T_\nu^2  f_1(y_i)
    + 2 T_\nu^4 f_2(y_i)
\right],
\label{eq:phiRel}
\end{align}
where 
$f_n(y_i) 
\equiv
  2  
  \int_{y_i}^\infty dy 
  y^{2n}
  d [e^{y} + 1]^{-1} / d y
= - \int_{y_i}^\infty dy y^{2n} 
(1 + \cosh y)^{-1}$. Note that the $\Theta$-function has
been converted to
$y \geq y_i$ with $y_i \equiv $ $m_i |{\bm v}_{\nu_i c} \cdot 
\hat {\bm k}| T_\nu^{-1} [1 - ({\bm v}_{\nu_i c} \cdot \hat {\bm k})^2]^{-1/2}$
and implemented in the integration.
The current bound on the neutrino mass 
sum is $\sum m_i < 0.13$\,eV at 95\% C.L. 
\cite{Workman:2022ynf,DES:2021wwk}, which translates to
$m_{\nu}^{\rm lightest} \lesssim 0.035$\,eV for NO. 
Then, the heaviest neutrino mass is 
$m_{\nu}^{\rm heaviest}\lesssim 0.07$\,eV 
and $y_i \lesssim 0.5$. 
The functions $f_1(y_i)$ and $f_2(y_i)$ 
can be safely approximated by their 
values at $y_i = 0$, $f_1(0) = -\pi^2/3$
and $f_2(0) = - 7\pi^4/15$, with a precision of $0.7\%$
since the next order is $\mathcal O (y^3_i)$. For $f_0$, the next order 
expansion, $f_0(y_i) \approx -1 + y_i/2$, 
is necessary to reach the $0.5\%$ precision.
Altogether, \geqn{eq:phiRel} is approximately,
\begin{align}
  \widetilde \phi 
\approx 
- \frac{ G a^2}{|\bm k|^2}
  \sum_i
  ({\bm v}_{\nu_i c} \cdot \hat {\bm k})
\left[ 
  m_i^4 
  \left(1 - \frac{m_i |{\bm v}_{\nu_i c} \cdot \hat {\bm k}|}{2T_\nu}
  \right)
+ \pi^2 m_i^2 T_\nu^2  
+\frac{14}{15} \pi^4 T_\nu^4 
\right].
\label{eq:TildephiRel}
\end{align}
Comparing with \cite{Okoli:2016vmd},
the relativistic corrections $m_i^2T_\nu^2$
and $T_\nu^4$ that are important for relativistic neutrinos
are also taken into account.
In addition, we keep the linear correction from
$m_i$ in the parenthesis whose effect can reach 25\%
for the heaviest neutrino.
The overall minus sign is very important since it can
determine whether the overdensity increases or decreases 
along the C$\nu$F velocity.

In principle, the relativistic components should also
include the CMB photon and the cosmic gravitational 
focusing can also happen between the CMB photons and 
DM halo.
However, the massless photons
can only contribute via the $T^4_\gamma$ term which is negligibly smaller than the C$\nu$F contribution. Even at redshift $z = 10$,
the CMB temperature is around only $10^{-3}$\,eV which is at least 
one order smaller than
$\sqrt{\Delta m^2_{13}} \approx 5 \times 10^{-2}$\,eV.
With the fourth power, the CMB contribution is suppressed by at
least four orders and can be safely omitted.
A similar effect occurs between 
neutrino halos and the background photon to produce
a dipole term in the CMB spectrum \cite{Kaplinghat:2003bh,Hotinli:2023scz}.

\subsubsection{Cosmic Gravitational Focusing with a Point Source}

Although the classical picture with single particle trajectory
provided in \gsec{sec:trajectory} can also handle the
density perturbation using the Liouville's theorem \cite{Alenazi:2006wu},
it is
much more convenient to use the Boltzmann equation formalism
which intrinsically handles overdensities. For easy comparison,
we apply the Boltzmann formalism to a point source of
gravitational potential.

Moving to the real space by Fourier transformation, 
the imaginary term of density fluctuation
$ \delta  \widetilde \rho_m({\bf k})
= \rho_m \widetilde \delta_{m 0} ( 1 + i \widetilde \phi )$
with $\widetilde \phi$ in \geqn{eq:TildephiRel} becomes,
\begin{align} 
  \delta \rho_m ({\bm x})
\propto
- i \int \frac{d^3 \bm k}{(2 \pi)^3} 
  e^{i \bm k \cdot \bm x}
  \frac 1 {|{\bm k}|^2}
  \tilde \delta_{m 0} (\bm k)
  (\bm v_{\nu c} \cdot \hat{\bm k}).
\label{eq:dRhoReal}
\end{align} 
For a point DM distribution at $\bm x_i$, i.e., 
$\delta_{m 0} (\bm x) \propto \sum_i \delta^{(3)}_D (\bm x - \bm x_i)$
in the real space, the DM overdensity is plane wave,
$\tilde \delta_{m 0} (\bm k) \propto \sum_i e^{ - i \bm k \cdot \bm x_i}$,
in the Fourier space.
Consequently, the Fourier transformation in \geqn{eq:dRhoReal}
can be carried out analytically,
\begin{align} 
  \delta \rho_m ({\bm x})
\propto
- i 
  \sum_i
  \int \frac{d |\bm k| }{(2 \pi)^3} 
  \int d \cos \theta
  e^{i |\bm k| |\bm x - \bm x_i| \cos \theta }
  \int (\bm v_{\nu c} \cdot \hat{\bm k}) d \phi.
\end{align} 
Choosing the relative position
$\bm r_i \equiv \bm x - \bm x_i$ as the $\bm z$ axis,
the other two vectors $\bm k$ and $\bm v_{\nu c}$ can be
parameterized as,
$\hat{\bm k} 
\equiv (\sin \theta \cos \phi, \sin \theta \sin \phi, \cos \theta)$ 
and 
$\hat{\bm v}_{\nu c} 
\equiv (\sin \theta_v \cos \phi_v, \sin \theta_v \sin \phi_v, \cos \theta_v)$. 
After integration, the $(x,y)$ components vanish
since $\int ( \cos \phi, \sin \phi ) d \phi = 0$ and
only the third component $\sim \cos \theta \cos \theta_v$ survives.
Then, we can directly integrate out the zenith angle
$\cos \theta$,
\begin{align} 
  \delta \rho_m ({\bm x})
\propto
  2 
  \sum_i
  (\bm v_{\nu c} \cdot \bm r_i)
  \int \frac{d |\bm k| }{(2 \pi)^2} 
  \frac{ J_1 (|\bm k| |\bm r_i|) }{  |\bm r_i|},
\label{eq:cosvwk}
\end{align} 
where $J_1 (x) \equiv \sin x/x^2 - \cos x / x$, 
is the first spherical Bessel function. We see that the  
$-i$ factor is absorpted into the $d \cos \theta$
integration to give a real number.
The minus sign in \geqn{eq:TildephiRel} is to ensure that
C$\nu$F focuses downstream.
The prefactor $\hat{\bm v} \cdot \hat{\bm r}$ indicates
that the density increases along the direction of
$\bm v_{\nu c}$.

\section{Galaxy Correlation with Cosmic Gravitational Focusing} 
\label{sec:Observable}

Since both neutrinos and DM are invisible, their cosmic
gravitational focusing and the resultant density fluctuations cannot
be observed directly. Contributing to the matter density
fluctuation, C$\nu$F can affect galaxy distribution. So 
the galaxy correlation functions can be used to trace
the cosmic gravitational focusing. In particular,
the C$\nu$F focusing effect can manifest itself in 
the galaxy dipole correlation function as imaginary
power spectrum in the Fourier space
\cite{Zhu:2013tma,Okoli:2016vmd} as well as
the galaxy weak lensing \cite{Zhu:2014qma}.
We derive the imaginary power spectrum 
with the redshift-space distortion (RSD) in
\gsec{sec:Cosobs} to show its dependence on the
neutrino masses. The signal-to-noise ratio (SNR)
for observation is summarized in \gsec{sec:SNR_gg}.

\subsection{Imaginary Galaxy Power Spectrum with RSD}
\label{sec:Cosobs}

In the late Universe $( z \lesssim 100)$, 
the baryon matter tightly follows the density fluctuations
contributed by mainly DM but also the C$\nu$F to form galaxies. 
Consequently, the galaxy distribution encodes 
the information of the cold DM ($\delta_c$) and
C$\nu$F ($\delta_\nu$) overdensities.
To be more exact, the galaxy overdensity is a linear
function of the DM and C$\nu$F overdensities with
different bias \cite{LoVerde:2014pxa},
\begin{align}
  \delta_{g\alpha} 
= 
  b_c^\alpha F_c \delta_c 
+ b_\nu^\alpha F_\nu \delta_\nu
\quad \Rightarrow \quad 
  \widetilde \delta_{g \alpha}
\approx
  \widetilde \delta_{m 0} 
\left( 
  b_c^\alpha 
+ i b_\nu^\alpha \widetilde \phi
\right).
\label{eq:delta_ga}
\end{align}
In the approximation, we have used the fact that the real
part is mainly contributed by DM, 
${\rm Re}[\delta_{g \alpha}] \approx b_c^\alpha \delta_{m 0}$, 
while the imaginary one comes from C$\nu$F,
${\rm Im}[\delta_{g \alpha}] 
= b_\nu^\alpha {\rm Im} [F_\nu \delta_\nu]
= i b_\nu^\alpha \delta_{m 0} \widetilde \phi$.
As defined in \geqn{eq:dmgeneral},
the total matter density is
$\widetilde \delta_m 
\equiv \widetilde \delta_{m0}(1 + i \widetilde \phi)$
which implies that $F_c \widetilde \delta_c = \widetilde \delta_{m0}$ and
$F_\nu \widetilde \delta_\nu = i \delta_{m0} \widetilde \phi$.
Note that $F_\nu \widetilde \delta_\nu$ here is essentially
its imaginary part while its real part can be
neglected or already combined into $F_c \widetilde \delta_c$.

The observed galaxy overdensity in the redshift space 
($\delta_{g \alpha, {\rm RSD}}$) is different from 
the one in the real space
($\delta_{g \alpha}$) 
due to the Doppler shift caused by the galaxy peculiar velocity $\bm u_m$.
This effect is known as the RSD \cite{Dodelson:cosmo2nd},
\begin{eqnarray}
  \delta_{g \alpha, {\rm RSD}} (\bm x)
\equiv
  \delta_{g \alpha} (\bm x)
- \frac{\partial}{\partial x} 
  \left( 
    \frac{\bm u_m \cdot \hat{\bm x}}{aH}
  \right).
\end{eqnarray}
For a specific Fourier mode with wavenumber $\bm k$,
the spatial derivative
$\partial/\partial x 
\equiv \hat{\bm x} \cdot \nabla_{\bm x}$ becomes
$\hat{\bm x} \cdot i \bm k$. Since the curl part
dampes out during the cosmic evolution \cite{Bernardeau:2001qr},
the peculiar velocity can be reparameterized,
$\widetilde{\bm u}_m = -i \widetilde \theta_m \bm k / |\bm k|^2$,
in terms of the velocity divergence
$\widetilde \theta_m$ which can be further expressed
in terms of $\dot{\widetilde \delta}_m$ according
to the linear Boltzmann equation,
$a \dot{\widetilde \delta}_m +  \widetilde \theta_{m} \approx 0$.
Altogether, the overdensity with RSD becomes,
$\widetilde  \delta_{g \alpha, {\rm RSD}}
=
  \widetilde \delta_{g \alpha}
+ \mu_{\bm k}^2 \dot {\widetilde \delta}_m / H
$
where $\mu_{\bm k} \equiv \hat{\bm x} \cdot \hat{\bm k}$.
Together with the $\widetilde \delta_{g \alpha}$ in
\geqn{eq:delta_ga} and the matter overdensity
$\widetilde \delta_m$ elaborated in \gsec{sec:deltam}, 
$\widetilde  \delta_{g \alpha, {\rm RSD}}$ becomes,
\begin{align} 
 \widetilde  \delta_{g \alpha, {\rm RSD}}
=
  b_\alpha^c \widetilde \delta_{m 0} 
+
  \frac{\mu_{\bm k}^2}  H  
   \dot{\widetilde \delta}_{m0}
+ i 
\left[  
  b^\nu_\alpha \widetilde \delta_{m 0} \widetilde \phi
+
  \frac{\mu_{\bm k}^2}  H  
  \left(
    \dot{\widetilde \delta}_{m0}
    \widetilde \phi
  +
    \widetilde \delta_{m0}
    \dot{\widetilde \phi}
  \right)
\right].
\end{align} 
For convenience, we have disentangled the real
and imaginary parts.

The observable in galaxy survey is the 
galaxy two-point cross correlation functions, 
$\mathcal P_{\alpha \beta} \equiv \widetilde \delta_{g \alpha, {\rm RSD}} \widetilde \delta_{g \beta, {\rm RSD}}^*$,
whose imaginary part is non-zero only if $\widetilde \phi \neq 0$.
For generality, we use $\alpha$ and $\beta$ to indicate
different galaxy categories for cross correlation.
The imaginary part of the galaxy cross correlation is,
\begin{align} 
{\rm Im}[\mathcal P_{\alpha \beta}]
= & 
  - ( b^\alpha_c b^\beta_\nu - b^\alpha_\nu b^\beta_c )
  \widetilde \delta_{m0}^2 \widetilde \phi
-
   (b^\alpha_c - b^\beta_c) 
  \frac{\mu_{\bm k}^2} H
  \widetilde \delta_{m0} 
  ( \dot{\widetilde \delta}_{m0} \widetilde \phi 
  + \widetilde \delta_{m0} \dot{\widetilde \phi})
+
  ( b^\alpha_\nu - b^\beta_\nu )
  \frac{\mu_{\bm k}^2} H
  \dot{\widetilde \delta}_{m0} \widetilde \delta_{m0} \widetilde \phi.
\label{eq:gg_general_bias}
\end{align}     
with general bias $b_c$ and $b_\nu$ for DM and
C$\nu$F components, respectively. The values of these
bias parameters are up to choice. For example,
Ref.\cite{Zhu:2013tma} takes $b_\nu^\alpha 
= b_\nu^\beta = 1$ but does not consider the RSD
correction while Ref.\cite{Okoli:2016vmd}
considers the RSD correction but assumes 
$b_\nu^{\alpha, \beta} = b_c^{\alpha, \beta}$.

Notice that the observable in \geqn{eq:gg_general_bias} is zero 
if both $b_c^\alpha = b_c^\beta$ and $b^\alpha_\nu = b^\beta_\nu$.
In other words, the cosmic gravitational focusing can be measured
only via the cross correlation of two galaxy categories that have
at least one different bias.
The neutrino bias is close to 1 up to $\lesssim 2\%$ 
corrections \cite{LoVerde:2014pxa}. For simplicity,
we take $b_\nu^\alpha = b_\nu^\beta  = 1$.
\geqn{eq:gg_general_bias} can be further simplified by
using $\dot{\widetilde \delta}_{m 0} = H f \widetilde \delta_{m0}$,
\begin{align} 
 {\rm Im}[
 \widetilde \delta_{g \alpha, {\rm RSD}} 
 \widetilde \delta_{g \beta, {\rm RSD}}^* ]  
= 
-
  i \Delta b
\left[ 
  \mu_{\bm k}^2  
  \frac{\dot{\widetilde \phi}} H  
  + (f \mu_{\bm k}^2 + 1 ) \widetilde \phi
\right]
 \widetilde \delta_{m0}^2,
\label{eq:Observable}
\end{align} 
where $\Delta b \equiv b_c^\alpha - b_c^\beta $
and $f \equiv d \ln D_+ /d \ln a \approx \Omega_m (z)^\gamma$
with $\gamma \approx 0.55$ \cite{DESI:2016fyo}.
Asumming $b_\nu^{\alpha, \beta} = b_c^{\alpha, \beta}$,
all those $\widetilde \phi$ terms in \geqn{eq:gg_general_bias}
vanishes and only the first term of \geqn{eq:Observable}
can survive \cite{Okoli:2016vmd}. Our result has the
additional second term with $b^\alpha_\nu \neq b^\alpha_c$
which is more reasonable since C$\nu$F and DM should have
different bias with different clustering properties
\cite{LoVerde:2014pxa}.
Since the two terms add up with a positive
sign in between, the different bias between neutrino and DM 
would amplify the cross correlation. This would make the
cosmic gravitational focusing more sensitive to the
neutrino masses.

\subsection{The Signal-to-Noise Ratio of Galaxy Cross-Correlation}
\label{sec:SNR_gg}

The cosmic gravitational focusing signal is defined 
as the imaginary part of the galaxy cross-correlation 
\cite{Okoli:2016vmd} as shown in \geqn{eq:Observable}, 
$\mathcal S \equiv {\rm Im}[ \widetilde \delta_{g\alpha, {\rm RSD}} 
\widetilde \delta^*_{g\beta, {\rm RSD}}]$. 
However, a typical cosmological observation is subject to two 
sources of statistical uncertainties: 
(1) The random fluctuations of the 
observable (in our case the galaxy overdensity 
$\delta_g$) and (2) A Poisson white noise ($\epsilon$). 
In the presence of a Poisson noise, 
the observed signal is shifted from the true observable, 
$ \widetilde \delta_{g\alpha} 
\rightarrow  {\widetilde \delta}_{g\alpha}' 
\equiv \widetilde \delta_{g\alpha} + \epsilon_{\alpha}$
 \cite{Dodelson:cosmo2nd}. Since $\delta_{g\alpha}$ 
 and $\epsilon_\alpha$ are independent, any 
dependence on $\epsilon_{\alpha}$ averages 
out for both linear and quadratic terms with $\alpha \neq \beta$.
Then the ensemble average of $\mathcal S\equiv {\rm Im}[ \widetilde \delta'_{g\alpha, {\rm RSD}} 
\widetilde \delta'^*_{g\beta, {\rm RSD}}]$ in the
presence of white noise,
 \begin{align}
  \langle \mathcal S \rangle 
\equiv
 {\rm Im}[ \langle \widetilde \delta'_{g\alpha, {\rm RSD}} 
  \widetilde \delta_{g\beta, {\rm RSD}}'^* \rangle ] 
=
 {\rm Im}[ \langle \widetilde \delta_{g\alpha, {\rm RSD}} 
  \widetilde \delta_{g\beta, {\rm RSD}}^* \rangle ]
=
 {\rm Im}[P_{\alpha \beta}],
 \label{eq:signal_def}
\end{align}
reduces back to the original imaginary parts of the galaxy 
power spectrum,
$P_{\alpha \beta} \equiv  
\langle \widetilde \delta_{g\alpha, {\rm RSD}} 
\widetilde \delta_{g\beta, {\rm RSD}}^* \rangle$.
In other words, the influence of white noises $\epsilon_\alpha$
diminishes in the cross correlation signal.

The noise/uncertainty of the measurement can be estimated
as the variance of the signal  
$  \mathcal N^2 
\equiv 
\left \langle 
  (\mathcal S - \langle \mathcal S\rangle )^2 
\right\rangle$ \cite{McDonald:2009ud}.
More concretely, it can be written
as the determinant of the covariance matrix $C_{\alpha \beta}$ 
in terms of the galaxy power spectrum 
and comoving number densities $n_\alpha(\equiv 
\langle \epsilon_\alpha \epsilon_\alpha^*\rangle^{-1})$ 
\cite{McDonald:2009ud},
\begin{align}
  \mathcal N^2 
=
  \frac 1 2 {\rm Det}[C]
+
  2 {\rm Im}^2[P_{\alpha \beta}]
\approx
  \frac 1 2 {\rm Det}[C]
\quad {\rm with} \quad 
  C_{\alpha \beta}
\equiv 
\left( 
   \begin{matrix}
      P_{\alpha \alpha} + n_\alpha^{-1} 
    &
      P_{\alpha \beta} 
    \\
      P_{\beta \alpha} 
    &
      P_{\beta \beta} + n_\beta^{-1} 
   \end{matrix}
\right).
\label{eq:variance}
\end{align}
Since ${\rm Im}[P_{\alpha \beta}] \ll {\rm Det}[C]$,
the error may be safely approximated using 
the half determinant, $\mathcal N^2 \approx {\rm Det}[C]/2$ 
\cite{Okoli:2016vmd,McDonald:2009ud}. 

We construct the log-likelihood 
$\langle \mathcal S \rangle^2 / \mathcal N^2$ 
from the SNR using \geqn{eq:Observable}
and \geqn{eq:variance}. The total sensitivity is obtained
by integrating over the wave number $\bm k$ and summing up
the redshift $z_i$ bins,
\begin{align}
  - 2\ln  L
=
  \sum_{z_i,\nu_j} 
  V_i
  \int \frac{d^3{\bm k}}{(2\pi)^3} 
  \frac{2\Delta b }{\rm Det[C]}
  \left \langle 
\left[ 
  \mu_{\bm k}^2  
  \frac{\dot{\widetilde \phi}_{\nu_j} } H  
  + (f \mu_{\bm k}^2 + 1 ) \widetilde \phi_{\nu_j}
\right]
  \widetilde \delta_{m0}^2
  \right \rangle^2,
\label{eq:LikelihoodDef}
\end{align}
where $V_i$ is the volume for the $z_i$ redshift bin.

Since
$  \widetilde \phi 
\propto 
  {\bm v}_{\nu c} 
= 
  {\bm v}_{\nu c}^{\rm bg} 
+ {\bm u}_{\nu c}$
which is a linear combination of the
background velocity ${\bm v}_{\nu c}^{\rm bg}$
and the relative velocity fluctuation ${\bm u}_{\nu c}$,
then 
$  \langle \widetilde \phi \widetilde \delta_{m 0}^2 \rangle 
\propto 
  {\bm v}_{\nu  c}^{\rm bg} 
  \langle \widetilde \delta^2_{m0}\rangle 
+ \langle {\bm u}_{\nu c} 
  \widetilde \delta^2_{m0}\rangle $ 
with 
$ \langle {\bm u}_{\nu c}\rangle = 0$. 
The second term is a third power of random Gaussian variables and is zero.
To calculate $ {\bm v}_{\nu c}^{\rm bg}$,
we make the approximation 
$  {\bm v}_{\nu c}^{\rm bg} 
= 
  \sqrt{\langle {\bm v}^2_{\nu c}\rangle} 
+ 
  \mathcal O({\bm u}_{\nu c}^2)$ 
\cite{Zhu:2013tma,Okoli:2016vmd,Tseliakhovich:2010bj,Yoo:2013qla}, thus 
$  \langle \widetilde \phi \widetilde \delta^2_{m0}\rangle 
\propto 
  \sqrt{\langle {\bm v}^2_{\nu  c}\rangle} 
  \langle \widetilde \delta^2_{m0} \rangle $.
In other words, 
$  \langle \tilde \phi \tilde \delta^2_{m0} \rangle 
\approx 
  \sqrt{\langle \widetilde \phi^2\rangle}P_{m 0}$, 
where $P_{m 0} \equiv \langle \delta_{m 0}^2 \rangle$
is the matter power spectrum seeded in the early Universe.
Similar arguments can also apply for $\dot{\widetilde \phi}$.
The ensemble average of $\widetilde \phi^2$ does not depend on the 
direction of the wave number $\bm k$ and the integration over 
$d\Omega_{\bm k}$ in \geqn{eq:LikelihoodDef} 
can be performed explicitly,
\begin{align} 
\hspace{-0.7cm}
  - 2\ln L
= &
  \sum_{z_i,\nu_j} 
  \frac{\Delta b^2 V_i}{5 \pi^2}
  \int d \ln |\bm k|
  \frac{|{\bm k}|^3 P_{m0}^2}{\rm Det[C]}
\left\{
  \left[
    \frac{\sqrt{\langle \dot{\widetilde \phi^2_i} \rangle}}H
  +\left(f + \frac{5}{3} \right)
    \sqrt{\langle \widetilde \phi^2\rangle}
  \right]^2
 + \frac{20}{9}
 \langle \widetilde \phi^2\rangle
\right\}.
\label{eq:lnl_dpdppp} 
\end{align}
Notice that the terms containing $\langle \widetilde \phi^2 \rangle$ 
in \geqn{eq:lnl_dpdppp} are extensions of the results in
\cite{Okoli:2016vmd}, especially the Eq.(A13) therein. Both
$\langle \widetilde \phi^2 \rangle$ terms are extra positive
contributions.
The ensemble average values 
$\langle \dot{\widetilde \phi^2_i} \rangle$ and 
$\langle \widetilde \phi^2\rangle$ can be found in \gapp{app:average}.
These averages are proportional to 
$\langle \bm v_{\nu c}^2 \rangle $, which decreases at large
scale according to the N-body simulation \cite{Inman:2016prk}.
We add a $\Theta$ function \cite{Okoli:2016vmd}  
inside the velocity integration to capture this feature,
as explicitly shown in \geqn{eq:vnuc_Theta}.

The imaginary contribution to the cross power spectrum
in \geqn{eq:LikelihoodDef} can be understood as a 
manifestation of the squeezed-limit bispectrum. 
When $| \bm k_1| \approx 0$,
the average
$ \langle v (\bm k_1) \delta_m (\bm k) \delta_m (\bm k') \rangle$
can be approximated as
$v \langle \delta_m^2 \rangle  $,
where the coefficient $v$ is the response
of the cross power spectrum to the presence of a long-wavelength relative
velocity field. 
This procedure is similar to the squeezed limits
of the nonlinear matter power spectrum
in the presence of a long-wavelength density field
\cite{Chiang:2014oga,Wagner:2014aka,Wagner:2015gva}.
In reality, the two-point functions,
viewed as squeezed limit of the three-point functions,
encapsulate the same information of
the gravitational focusing effect as the latter.

\section{Galaxy Number Density and Bias at DESI}
\label{sec:DESEresult}

The Dark Energy Spectroscopic Instrument (DESI) 
\cite{DESI:2016fyo,DESI:2018ymu} is a ground-based experiment 
to collect around 40 million galaxies and quasars in 5 years
by covering 14,000 square degree of the sky. 
The DESI instrument, the Mayall Telescope at the 
Kitt Peak National Observatory, is capable of 
detecting light from extragalactic sources in the
three optical bands ($r$, $g$, and $b$) with wave 
length from 360 nm to 980 nm.  
The DESI survey is expected to detect around 10 million bright 
galaxies for low redshift ($z < 0.6$) and 
30 million faint galaxies and quasars at intermediate 
redshift ($0.6 < z < 2$). The observation and 
characterization capabilities of the DESI experiments 
are perfect for detecting the cosmic gravitational
focusing effect. So we take DESI for illustration.
More specifically, we use halo occupation distribution
(HOD) based on the lightcone catalogue
\cite{Smith:2017tzz,Smith:2022hpr}
to determine the BGS number density in \gsec{sec:ngbg}
and bias in \gsec{sec:BGS-bias}. Finally, the faint
galaxies are summarized in \gsec{sec:faint}.

\subsection{BGS Number Density}
\label{sec:ngbg}

As pointed out above \geqn{eq:Observable} that the cosmic
gravitational focusing
appears in the cross correlation function of different galaxy
categories. The DESI instrument can make spectroscopic
observations of four distinct categories of extragalactic
sources \cite{DESI:2016fyo}:
the bright galaxy sample (BGS) at $ 0.05< z < 0.4$, 
luminous red galaxies (LRGs) at $0.4 < z < 1.0$,
star-forming emission line galaxies (ELG) at $0.6 < z < 1.6$,
and quasi-stellar objects (QSO)  
at $z > 0.6$. 
In the low redshift region $z<0.4$, BGS has the largest
population. To maximize the SNR, we split the BGS
category into two and consider the internal cross correlation.
The remaining LRG, ELG, and QSO can also contribute
with three cross correlations among them to be discussed
in \gsec{sec:faint}.

\begin{figure}[t]
  \centering
  \includegraphics[scale=0.5]{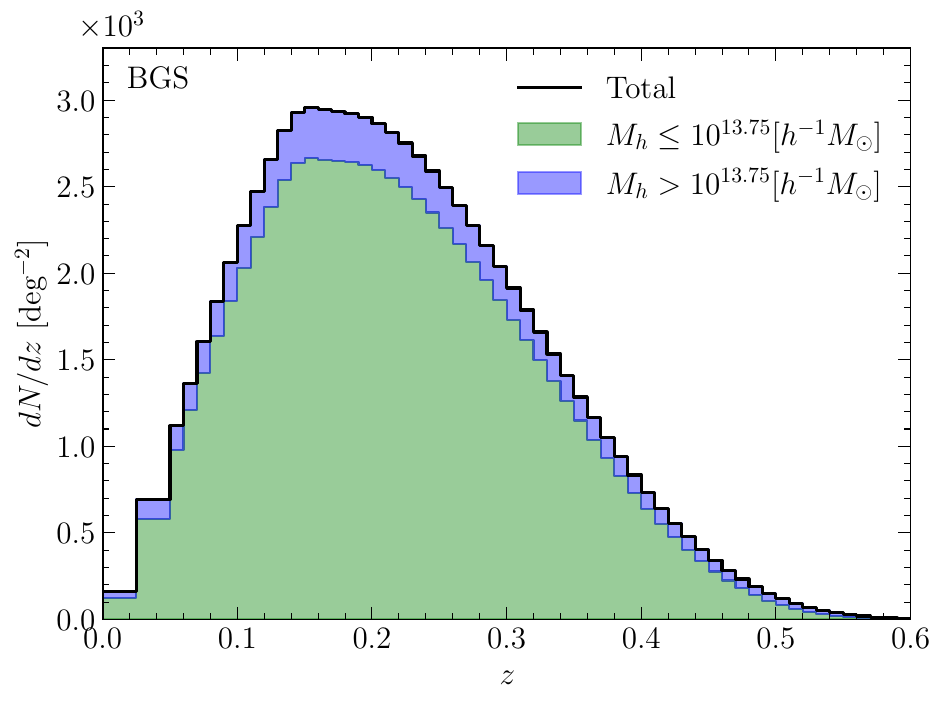}
  \includegraphics[scale=0.485]{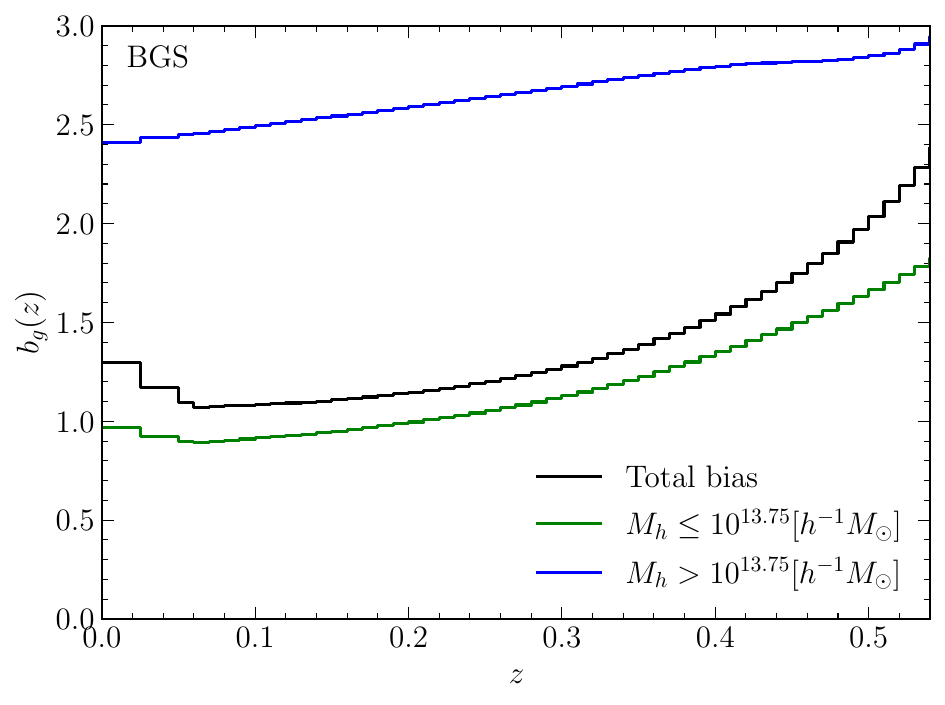}
\caption{The observable galaxy number distribution
(Left) and bias (Right)
for the BGS category by using the lightcone
catalogue \cite{Smith:2017tzz,Smith:2022hpr}. The
two sub-categories are divided according to the
halo mass $M_h < 10^{13.75} h^{-1} M_\odot$ (green)
and $M_h > 10^{13.75} h^{-1} M_\odot$ (blue).
For comparison, the combined values are shown
as black lines in both panels.}
\label{fig:galaxy_catalogue}
\end{figure}

The galaxy formation is mainly determined by the halo
distribution, namely, the halo mass function (HMF)
$d n / d \ln M_h$ that describes the halo number density
$n$ as a function of the halo mass $M_h$
\cite{Press:1973iz,Mo:2010,Dodelson:cosmo2nd}.
Given a halo mass $M_h$, the HOD gives the average
number $\langle N(M_h) \rangle$ of galaxies \cite{Kravtsov:2003sg}.
Then their convolution gives the galaxy number density,
\begin{align}
  n_g (z) 
\equiv
  \int d\ln M_h
  \frac{d n(z)}{d\ln M_h}
  \langle N(M_h) \rangle .
\label{eq:ng}
\end{align}
The redshift dependence comes from the HMF $dn(z)/d\ln M_h$
which can be calculated by the 
\href{https://github.com/halomod/halomod}{halomod} 
package 
\cite{Murray:2013qza,Murray:2020dcd} with the 
SMT model configuration \cite{Sheth:1999su} 
and the cosmological parameters given by Planck 2018 
\cite{Planck:2018vyg}. The HOD
is calculated by
\href{https://github.com/amjsmith/hodpy}{hodpy} 
\cite{Smith:2017tzz,Smith:2022hpr} which uses a Gaussian 
empirical parametrization of $\langle N(M_h) \rangle $
\cite{SDSS:2010acc} to fit the Schechter 
luminosity function \cite{Schechter:1976} obtained 
from the GAMMA \cite{GAMA:2011} and SDSS 
\cite{SDSS:2002vxn} data (for a more detailed 
description of the method see \cite{Smith:2017tzz}).
Since only those halos with mass above a typical threshold
value $M_{\rm min} = 10^{11} h^{-1} M_\odot$ can host galaxies,
the HOD $\langle N (M_h) \rangle $ goes to zero for
$M_h < M_{\rm min}$. On the other hand, the heavy halo
abundance is exponentially suppressed. Above
$M_{\rm max} = 10^{16} h^{-1} M_\odot$ there is almost
no halo which can be imposed by requiring
$M_h < M_{\rm max}$.  

Finally, the total number of galaxies,
$d^2 N / d z d \Omega$, is obtained 
by multiplying galaxy number density $n_g$ by the
comoving volume per redshift and square degree
$d^2 V / d z d \Omega = \left[\int dz'/H(z')\right]^2$ 
$\left( \pi / 180 \right)^2$ $[\Delta z/H(z)]$.
The left panel of \gfig{fig:galaxy_catalogue}
shows the BGS distributions per square degree
for the two sub-categories,
$M_h < 10^{13.75} h^{-1} M_\odot$ (green) and
$M_h > 10^{13.75} h^{-1} M_\odot$ (blue).
Dividing galaxies into sub-categories according to
the halo mass \cite{Ginzburg:2019xsj} is adopted to
differentiate the halo bias as elaborated below in
\gsec{sec:BGS-bias}.
The total height of the two regions shows the 
total number of observable galaxies for the BGS
category. Since the number density $n_g$ decreases
while the volume increases with the redshift $z$, 
$\propto (1+z)^3$,
the galaxies number per square degree first
increases to reach its maximum at $z \approx 0.15$
and then decreases fast. In other words, the
most sensitivity contributed by the BGS category
should come from $z \in (0.1 \sim 0.3)$.

\subsection{BGS Bias}
\label{sec:BGS-bias}

The galaxy bias is obtained by integrating over the same
distribution in \geqn{eq:ng} but with the halo bias $b_h(M_h, z)$
\cite{Kravtsov:2003sg} as an extra weight,
\begin{align}
  b_g(z)
\equiv
  \frac{1}{n_g }
  \int
  d\ln M_h
  \frac{d n(z)}{d\ln M_h}
  \langle N(M_h) \rangle
  b_h(M_h, z).
\label{eq:b_galaxy}
\end{align}
In addition to the HMF, the redshift dependence also
comes from the halo bias function $b_h(M_h, z)$ for
which we use the ST99 \cite{Sheth:1999mn} configuration
in the 
\href{https://github.com/halomod/halomod}{halomod} 
package 
\cite{Murray:2013qza,Murray:2020dcd}. 

As emphasized above, the cosmic gravitational focusing
signal in \geqn{eq:Observable}
requires the separation of galaxies into two categories with
different bias. This can be realized by dividing the BGS
category into two sub-categories according to the halo mass,
$M_h < M_{\rm split}$ and $M_h > M_{\rm split}$ \cite{Ginzburg:2019xsj}. 
For an observation,
the halo mass can be related to its galaxy luminosity
\cite{Vale:2004yt} with a cluster finder algorithm
\cite{Yang:2004qi,Yang:2007yr} to reach a precision
of $\Delta \log_{10} M_h/M_\odot h^{-1} \sim 0.4$
\cite{Yang:2020eeb}.
Accordingly, the number densities $n_1$ ($n_2$) in \geqn{eq:ng}
are obtained
by integrating the halo mass $M_h$ over the two mass ranges,
$M_{\rm min} < M_h < M_{\rm split}$ and
$M_{\rm split} < M_h < M_{\rm max}$, respectively.

\begin{figure}[t]
  \centering
  \includegraphics[width=0.45\textwidth]{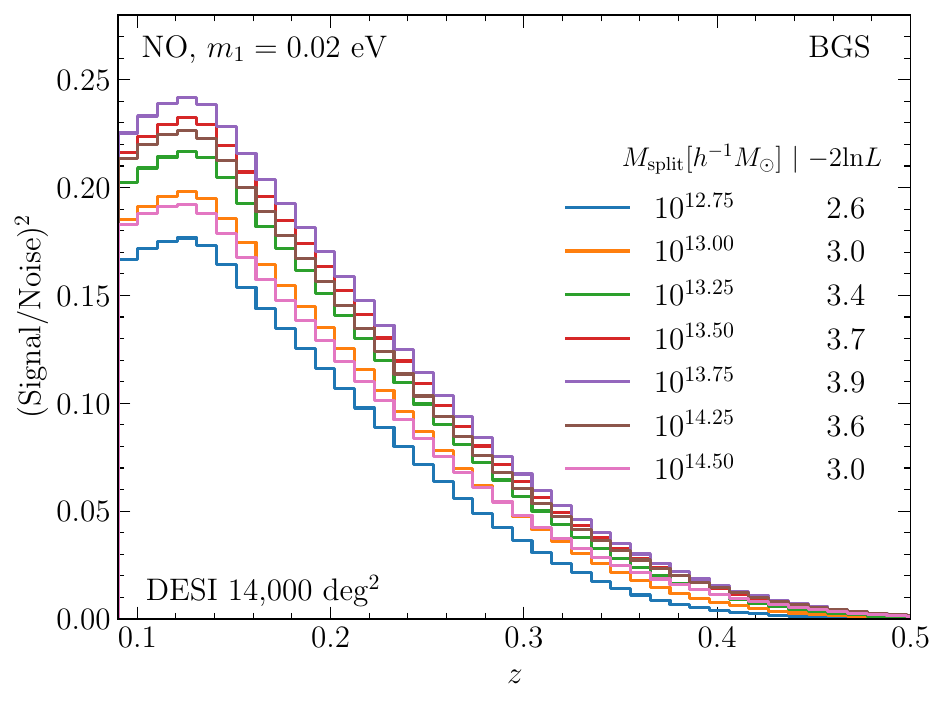}
  \includegraphics[width=0.45\textwidth]{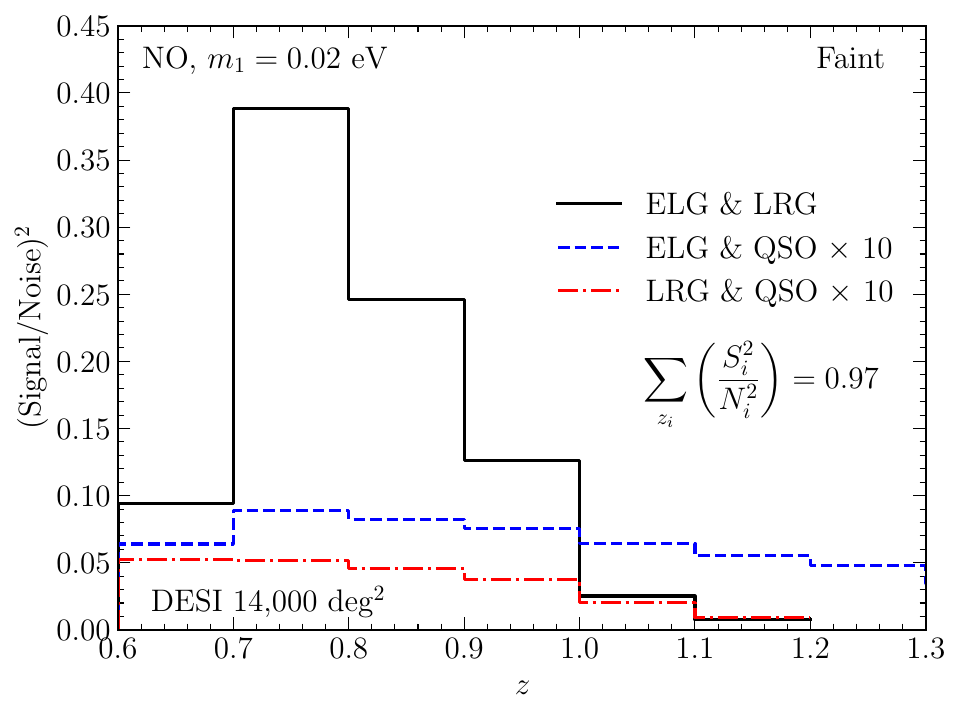}
\caption{
  The SNR of cosmic gravitational focusing for the BGS (Left)
  and remaining (Right) catalogues with the lightest
  neutrino mass $m_1 = 0.02$\,eV and the NO.
  While the BGS survey is mainly at the low redshift
  region $z<0.6$, the
  faint galaxy (LRG, ELG, and QSO) surveys reside at
  higher redshift $z>0.6$. A total area of 14,000 square
  degree can be observed at the DESI instrument.
}
\label{fig:sensitivity_CnuB}
\end{figure}

The SNR in \geqn{eq:lnl_dpdppp} is proportional to the bias 
difference of two galaxies
$\Delta b \equiv b_1 - b_2$. On the other hand, the 
shot noise decreases with $n_1$ and $n_2$ in the determinant of 
the covariance matrix according to \geqn{eq:variance}. Hence, 
the sensitivity increases with both bias difference $\Delta b$
and galaxy numbers $n_i$. As we have seen above, the 
mass splitting parameter $M_{\rm split}$ affects both $n_i$ and 
$\Delta b$ via integration limits.
Namely, if $M_{\rm split}$ increases, $\Delta b$
and $n_1$ increases while $n_2$ decreases, and vice versa. 
Although larger $\Delta b$ and $n_1$ can increase
the sensitivity, the decrease of $n_2$ will make it worse.
The optimal value of $M_{\rm split}$ can be found by
maximizing the SNR. Our result is shown in the left panel of
\gfig{fig:sensitivity_CnuB} for NO and $m_1 = 
0.02$\,eV. Several typical values in the range of
$M_{\rm split} \in [10^{12.75}, 10^{14.5}] h^{-1}M_\odot$
are shown for comparison. The total log-likelihood starts from $2.6$
with $M_{\rm split} = 10^{12.75} h^{-1}M_\odot$ and increases to 
its maximal value 3.9 with $M_{\rm split} = 10^{13.75} h^{-1}M_\odot$.
So we take $M_{\rm split} = 10^{13.75} h^{-1}M_\odot$
as a benchmark point for our results in \gfig{fig:galaxy_catalogue}
and later discussions.

The right panel of \gfig{fig:galaxy_catalogue} shows 
the galaxy bias of the BGS category. For the whole
BGS category, a major feature of the black line is that
the BGS bias increases with redshift. This is because a
flux-limited sample at high redshift is intrinsically
brighter and hence has a larger halo mass which usually
leads to more galaxies.
The green and blue lines denote the two sub-categories of the BGS:
$M_h < 10^{13.75} h^{-1} M_\odot$ and
$M_h > 10^{13.75} h^{-1} M_\odot$. The heavier sub-categories
in blue easily host more galaxies
and therefore have a larger bias $b_g >2.4$ than the
lighter sub-category one $2 > b_g > 1$.
The bias difference between the two sub-categories
approaches to $\Delta b \approx 1$ which is quite
sizable.

\begin{figure}[t]
\centering
\includegraphics[width=0.49\textwidth]{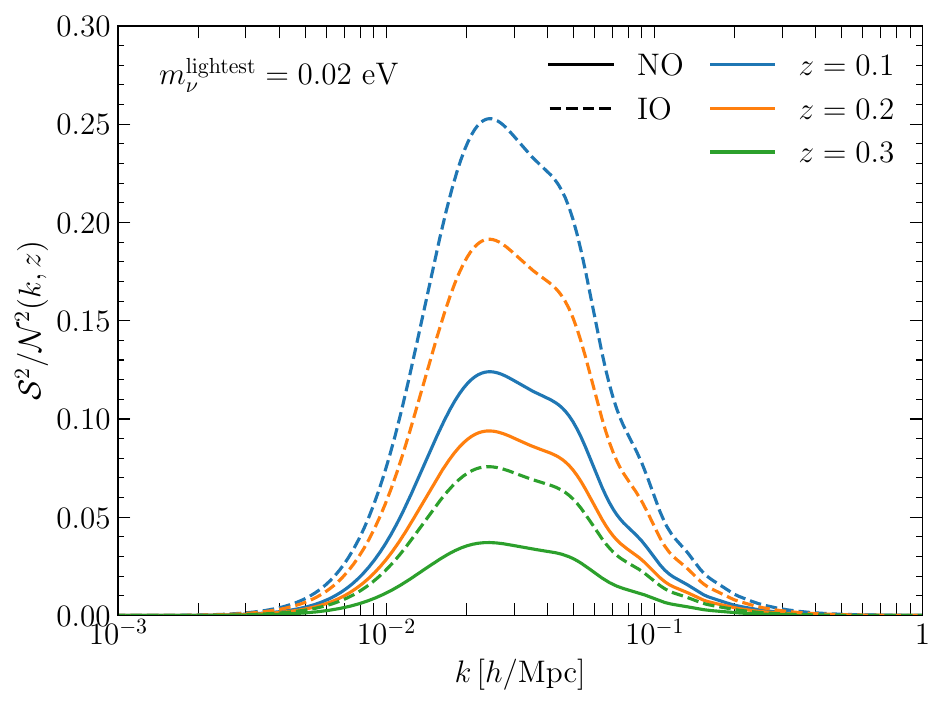}
\includegraphics[width=0.49\textwidth]{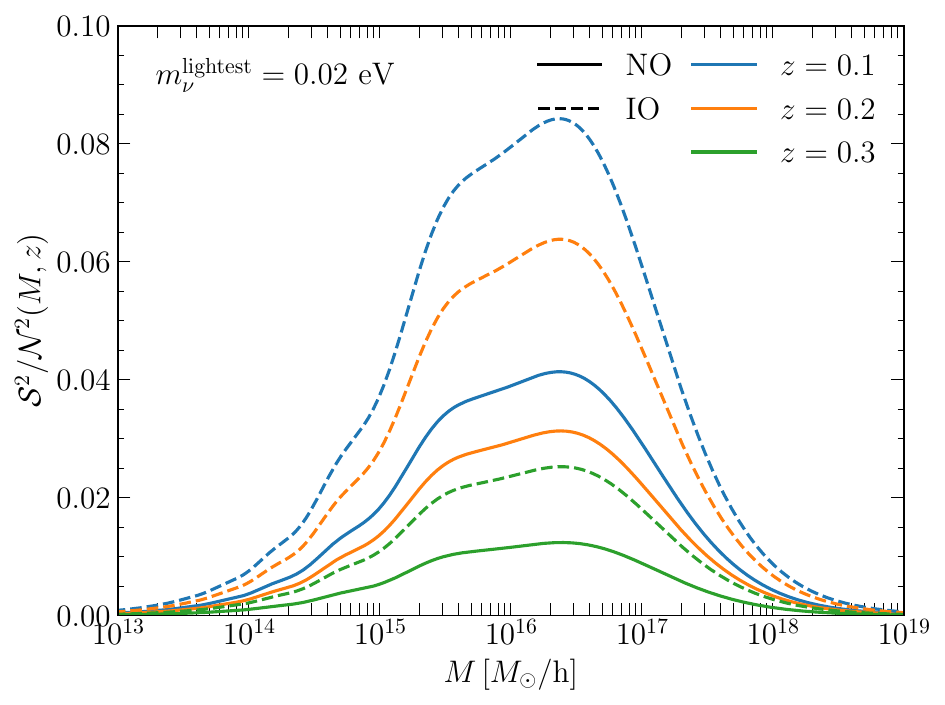}
\caption{
The SNR integrand for both NO and IO with the lightest
neutrino mass $m_\nu^{\rm lightest} = 0.02\,{\rm eV}$
at redshifts $z=0.1$, $0.2$, and $0.3$.
The left panel shows the SNR integrand as a function of
the comoving wave number $|\bm k|$ while the right panel
takes the mass enclosed
in the corresponding radius $R \equiv 1 / |\bm k|$.
}
\label{fig:SNR_kmode}
\end{figure}

\gfig{fig:SNR_kmode} shows the scales that contribute
the most to the SNR with three typical redshift
values $z = (0.1$, $0.2$, and $0.3$) for the BGS
sub-categories. What we plot is the integrand
in \geqn{eq:lnl_dpdppp} which is a function of the wave
number $\bm k$ and proportional to 
$|\bm k|^3 P_m^2(\bm k) \widetilde \phi^2$. 
With higher redshift, the contribution to SNR is smaller
since the matter clustering is smaller at ealier stages.
Between NO and IO, the latter has larger effect with two
heavy neutrinos. At large scales
(small $|\bm k|$), the integration kernel
$|\bm k|^3 P_m^2(\bm k) \widetilde \phi^2 \propto |\bm k|^3$
(with matter power spectrum $P_m(\bm k) \sim |\bm k|$ \cite{Dodelson:cosmo2nd}
and $\widetilde \phi \propto |\bm v| / |\bm k|^2 \propto  1 / |\bm k|$ 
in \geqn{eq:phiRel} 
since the velocity $\bm v \propto \nabla \Psi \propto |\bm k|$)
shown in \gfig{fig:SNR_kmode} is suppressed by small $\bm k$.
Also, the peak shifts from the $0.01\,h/{\rm Mpc}$ for
the power spectrum to $0.02\,h/{\rm Mpc}$.
The largest contribution to the SNR values occurs 
in the region of $10^{-2} h/{\rm Mpc} < k < 10^{-1} h/{\rm Mpc}$
which corresponds to the 
length scales of $(15 \sim 150)\,{\rm Mpc}$
with the dimensionless Hubble rate $h = 0.67$. 
This range
is inside the region where the linear bias choice in
\geqn{eq:delta_ga} applies \cite{Desjacques:2016bnm}.
Such linear bias does not suffer much from
the uncertainty at small scales comparing with the usual 
mass sum constraint based on the small scale suppression.
The decreasing on the large-$k$ end
comes from the matter power spectrum
$|\bm k|^3 P_m^2(\bm k) \widetilde \phi^2 
\sim 1/|{\bm k}|^5$ with $P_m (\bm k) \sim 1 / |\bm k|^3$
\cite{Dodelson:cosmo2nd}.

The integration over $|\bm k|$ in \geqn{eq:lnl_dpdppp}
can be rewritten,
$
  - 2 \ln L 
\equiv 
  \sum_{z, \nu_i} 
  \int   
  \mathcal S^2 / \mathcal N^2( |\bm k| , z) 
  d \ln |\bm k|
\equiv 
  \sum_{z, \nu_i} 
  \int 
  \mathcal S^2 / \mathcal N^2( M , z) 
  d \ln M
$,
in terms of the corresponding mass
$ M(R) \equiv \rho_m 4 \pi R^3 / 3 $
enclosed in the sphere with radius $R \equiv 1/|\bm k|$. 
Correspondingly, the integration interval changes as
$
  d M  
= 
- \rho_m 4 \pi   
  d |\bm k| / |\bm k|^4
= 
  - 3 M 
  d |\bm k| / | \bm k |
$
and consequently 
$
  d \ln M 
= 
  - 3 d \ln |\bm k|
$.
For large mass (small $|\bm k|$), the  
$\mathcal S^2/ \mathcal N^2 (M, z) \propto 1/M $
while for small mass (large $|\bm k|$), 
the $\mathcal S^2/ \mathcal N^2 ( M , z) \propto M^{5/3}$
as shown in the right panel of \gfig{fig:SNR_kmode}.
With the critical density 
$\rho_{\rm cr} = 2.7 \times 10^{11} h^2 M_\odot/ {\rm Mpc}^3$ and matter fraction
$\Omega_m \approx 0.3 $,
the corresponding mass region enclosed in the sphere
with radius $R = 1/|{\bm k}|$ ranges from
$10^{14} M_\odot/h$ to $10^{18} M_\odot/h$.
The peak radius $\sim 75\,{\rm Mpc}$ 
($|{\bm k}| \approx 0.02\,h/{\rm Mpc}$) in the left panel
corresponds to the peak mass $\sim 10^{16} M_\odot/h$
in the right panel. Since such a volume can accommodate
a few hundreds of halos with typical halo size of a few Mpc
\cite{Dodelson:cosmo2nd}, the corresponding halo mass
is consistent with our mass splitting point
$M_{\rm split} \times 100 \approx 10^{16} M_\odot/h$.
Between the two neutrino mass orderings,
IO (solid) has almost two times larger SNR
than NO (dashed) in both panels since the former
has two nearly degenerate heavy neutrinos
while later has only one.

\subsection{Faint Galaxies}
\label{sec:faint}

For the faint galaxy population at higher redshift
($z > 0.6$), including LRG, ELG, and QSO,
we use the number density distributions and
biases summarized in the DESI paper
\cite{DESI:2016fyo,DESI:2018ymu}, such as the Table 2.3
of \cite{DESI:2016fyo}. 
The galaxy bias is taken as 
$b_{\rm LRG} (z) D(z) = 1.7$ (LRG) \cite{DESI:2016fyo}, 
$b_{\rm ELG} (z) D(z) = 0.84$ (ELG) \cite{Mostek_2013},
and
$b_{\rm QSO}(z) D(z) = 1.2$ (QSO) \cite{Ross_2009},
respectively.
With dependence on the linear growth factor $D(z)$,
the galaxy bias is also a function of redshift.
In the right panel of \gfig{fig:sensitivity_CnuB},
we show the SNR for all the three possible correlations
among the faint galaxy categories. The total SNR is
roughly $0.97$ which is smaller than the BGS. 
Among them, the correlation between ELG and LRG dominates
which is a direct consequence 
of the fact that the observable QSO population is
much smaller than ELG and LRG.

\section{DESI Sensitivity and Synergy with Mass Sum Constraints}
\label{sec:Numass}

\begin{figure}[t]
\centering
\includegraphics[width=0.49\textwidth]{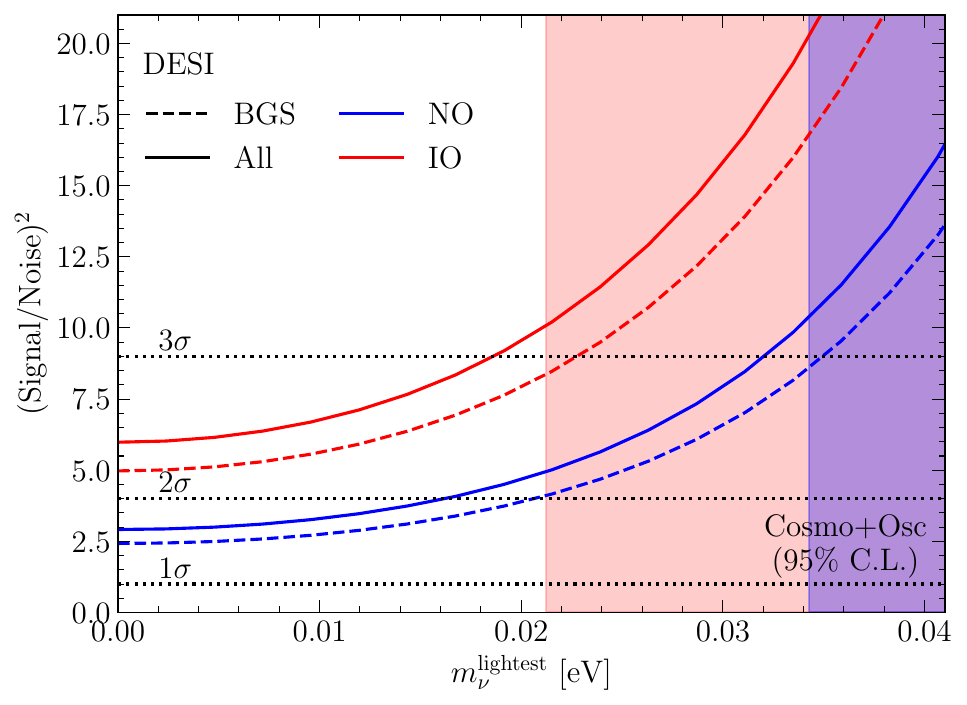}
\hfill
\includegraphics[width=0.49\textwidth] {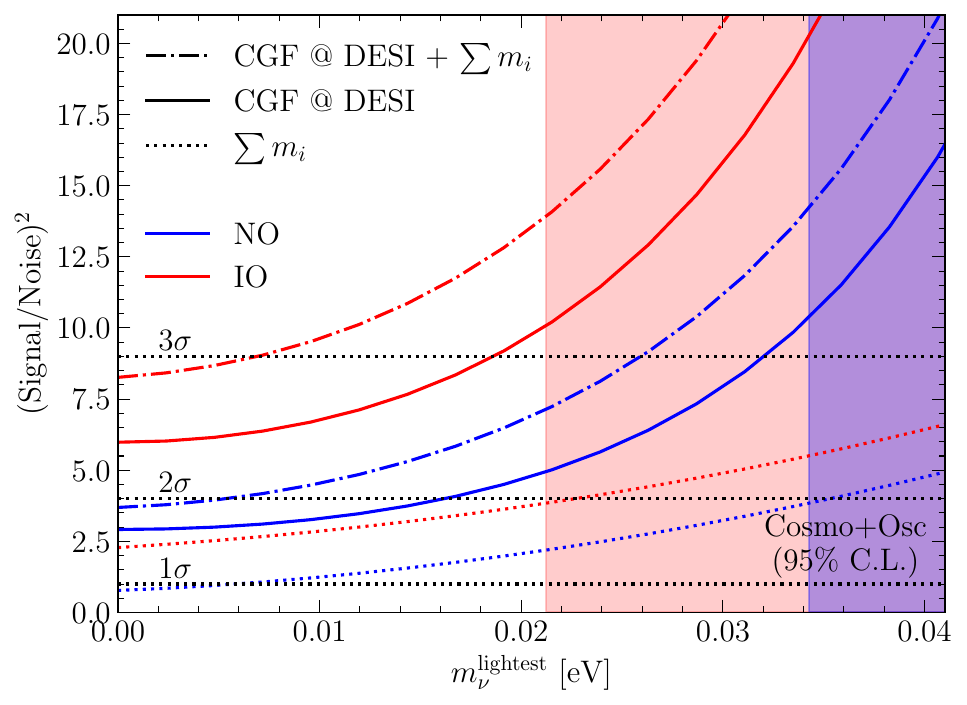}
\caption{
(Left) The SNR at DESI for both NO (blue) and IO (red)
cases as function of the lightest neutrino mass
($m^{\rm lightest}_\nu$) with BGS alone (dashed) or
all categories (solid). (Right)
The comparison of cosmic gravitational focusing
sensitivity at DESI (CGF @ DESI, solid) with the existing
cosmological constraints (dotted). Their synergy is also
plotted as dot-dashed line.
For both panels, the shaded regions indicate the current
constraints ($\sum m_i < 0.13 ~{\rm eV}$ at 95\% C.L.)
from cosmology and neutrino oscillation experiments for
both NO (blue) and IO (red).}
\label{fig:SNR_total}
\end{figure}

With the knowledge of galaxy distributions and bias,
we can estimate the DESI sensitivity on the lightest
neutrino mass. As already implemented in
\gfig{fig:sensitivity_CnuB}, the DESI instrument can
observe 14,000 square degrees. Putting the galaxy distribution
\geqn{eq:ng} and bias \geqn{eq:b_galaxy}
plotted in \gfig{fig:galaxy_catalogue} 
back into the log-likelihood estimator in \geqn{eq:LikelihoodDef},
we obtain the total SNR as a function of 
the lightest neutrino mass $m^{\rm lightest}_\nu \equiv m_1$
(NO) and $m^{\rm lightest}_\nu \equiv m_3$ (IO)
as blue and red lines, respectively, in \gfig{fig:SNR_total}.
In our calculations, we take the neutrino mass squared differences to be 
$\Delta m_{21}^2 = 7.54\times 10^{-5}$\,eV$^{2}$, 
$\Delta m_{31}^2 = 2.47\times 10^{-3}$\,eV$^{2}$ 
\cite{Workman:2022ynf}.

The left panel of \gfig{fig:SNR_total} displays the SNR
varying with the lightest neutrino mass $m^{\rm lightest}_\nu$,
for the BGS only
(dashed) and the combination of all categories (solid).
For comparison, the current constraint
($\sum m_i < 0.13 ~{\rm eV}$ at 95\% C.L.) by 
combining both cosmology \cite{Planck:2018vyg} and
neutrino oscillation \cite{Workman:2022ynf} has
been shown as shaded regions for both NO (blue)
and IO (red). We can see that the projected sensitivity
of cosmic gravitational focusing at DESI with only BGS
(blue dashed)
can already significantly exceed the existing one.
After including
the faint galaxies (solid), the total SNR can further
increase by 20\%.
The 95\% upper limit on the lightest neutrino mass can
reach $m_\nu^{\rm lightest} = m_1 < 0.016\,{\rm eV}$
for NO while the current one is 0.034\,eV. For IO,
the vanishing lightest mass is even beyond 2$\sigma$.
In other words, the IO scenario with vanishing lightest
mass is possible to be excluded with cosmic gravitational
focusing by more than 2$\sigma$ at DESI while the
current constraint can only reach around 0.02\,eV.
Including the cosmic gravitational focusing as a third
independent cosmological constraint, in addition to CMB
and LSS, on the neutrino mass is of profound significance.

From NO (blue) to IO (red), the SNR increases by
roughly a factor of 2 since IO has two nearly degenerate
heavy neutrino masses while NO has only one heavy mass.
As emphasized in the previous sections, the gravitational
focusing effect has fourth power dependence on the neutrino
masses. Heavy neutrinos leaves much more significant
effect in the cosmic gravitational focusing.

The right panel of \gfig{fig:SNR_total} shows the synergy
of the cosmic gravitational focusing and the current constraints.
We can see that the left side of the shaded regions
corresponds to the 95\%\,C.L. value of the dotted lines.
After combination with the projected sensitivity of DESI
gravitational focusing observation (dot-dashed), the vanishing
lightest mass scenario with IO (red) can be excluded up
to almost $3 \sigma$. Even for the case of NO (blue),
the upper limit can further enhance to
$m_{\rm lightest} < 3.2\,{\rm meV}$ at 95\%\,C.L.

\begin{figure}[t]
\centering
\includegraphics[width=0.49\textwidth]{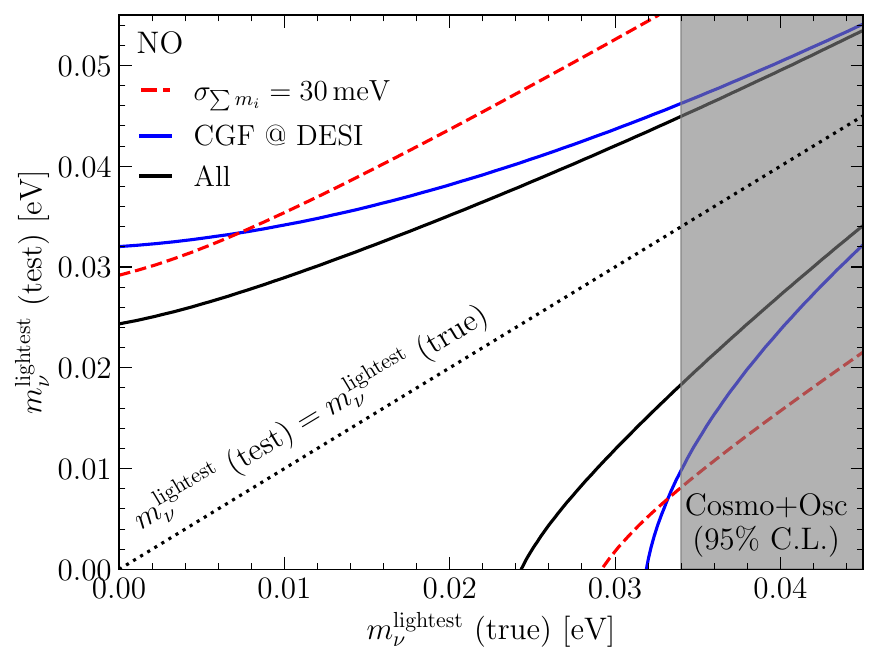}
\includegraphics[width=0.49\textwidth]{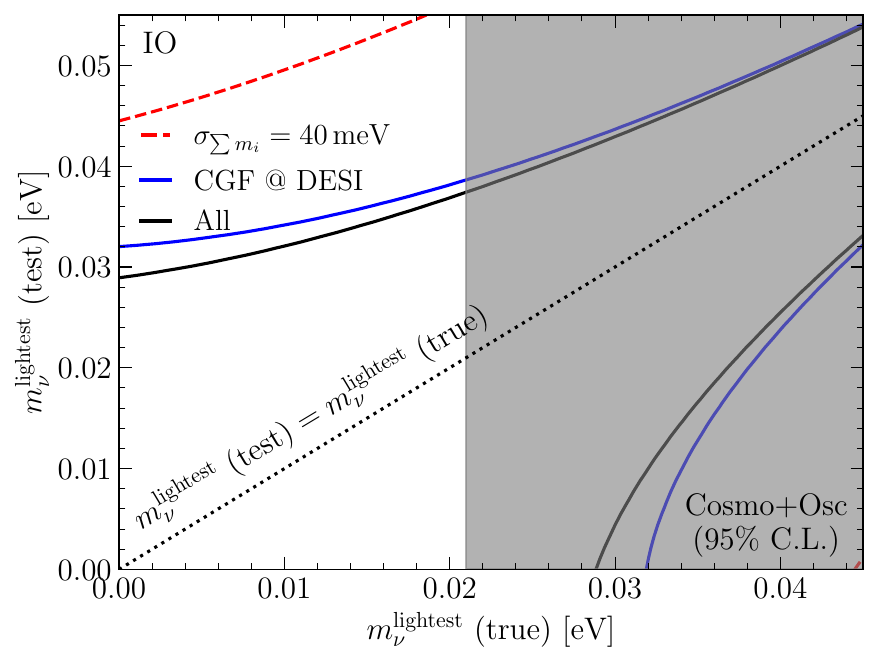}
\caption{
The upper and lower limits on the lightest neutrino
mass for NO (left) and IO (right). While the black
dotted line stands for the central value,
namely the test value being the same as the true one,
the lines above it indicate the upper limits and
those below the lower limits at 95\%\,C.L.. The future
cosmological observations ($\sigma_{\sum m_i} =
30 ~ (40)\,{\rm meV}$ for NO (IO) from CMB and LSS
\cite{Amendola:2016saw}) are shown
as red dashed lines, the projected DESI sensitivity
with cosmic gravitational focusing (CGF) as blue solid lines, and
their combination as black solid lines. For comparison,
the combination of existing cosmoligical constraints
and neutrino oscillation experiments at 95\%\,C.L.
\cite{Workman:2022ynf}
are shown with grey filled region.
}
\label{fig:error_SNR}
\end{figure}

Future galaxy surveys will further improve the
existing CMB and LSS constraints on the neutrino mass
sum. For example, the CSST
\cite{Zhan:2011,Cao:2018,Gong:2019yxt, Chen:2022wll}
will be able to reach
$\sigma_{\sum m_i} \sim 0.23$\,eV \cite{Lin:2022aro}.
In addition, DESI-II \cite{DESI:2022lza} and Subaru PFS
\cite{PFS_talk} can reach
$\sigma_{\sum m_i} \sim 0.02$\,eV, and
$\sigma_{\sum m_i} \sim 0.03$\,eV at EUCLID \cite{Amendola:2016saw} for NO,
in combination with the Plank 2018 data.
For the adopted mass sum, $\sum m_i = 0.125$\,eV,
the  combination of future LSS with Planck 2018 will reach 
$\sigma_{\sum m_i} = 0.03\,{\rm eV}$ for NO ($m_3 = 0.125\,{\rm eV}, m_1 \approx m_2 \approx 0$) and 
$\sigma_{\sum m_i} = 0.04\,{\rm eV}$ for IO ($m_1 \approx m_2 \approx 0.0625\,{\rm eV} , m_3 \approx 0$)
(See Table 5 of \cite{Carbone_2011}).
It is interesting to notice that
the combination of CMB and LSS observations has better
sensitivity for NO than IO due to the free-streaming
effect \cite{Lesgourgues:2004ps}. For the mass sum
$\sum m_i = 0.125$\,eV adopted above, the heaviest neutrino
mass ($m_3 = 0.125\,{\rm eV}$) for NO is nearly twice the
one ($m_1 \approx m_2 \approx 0.0625\,{\rm eV}$) for IO.
With a larger mass, the NO neutrino becomes non-relativistic
earlier and suppresses the power spectrum at small scales.
Consequently, the NO constraint is stronger than the IO one.
Note that these experimental sensitivities are
estimated for the smallest neutrino mass scenario
with $m_\nu^{\rm lightest} = 0$. For nonzero
lightest neutrino mass and hence larger mass sum
to be measured, its uncertainty $\sigma_{\sum m_i}$
is expected to become smaller with a larger observable.
As a conservative comparison, we take these nominal
values of projected uncertainties universally for a tunable
lightest neutrino mass to make
illustration in \gfig{fig:error_SNR}.
A perfect observation falls on the black dottd line 
$m^{\rm lightest}_\nu ({\rm test}) = m^{\rm lightest}_\nu ({\rm true}) $ while the distance from this line
represents the 95\%\,C.L. uncertainty for a given
$m_\nu ({\rm true})$.
Larger distance means larger uncertainty.
We can see that the other cosmological constraints
(red dashed) still span a wide range around the true
values (black dotted) even after taking the projected
sensitivities from future observations. This is
especially true for the IO (right) as well as the
NO (left) in most of the parameter space.

For the cosmic gravitational focusing constraint (blue solid)
in \gfig{fig:error_SNR}, its sensitivity is estimated
by replacing the signal strength in
\geqn{eq:signal_def} with the signal difference,
$\mathcal S \rightarrow
\mathcal S_{\rm true} - \mathcal S_{\rm test}$,
where $\mathcal S_{\rm true}$ ($\mathcal S_{\rm test}$)
is obtained with the true (test) neutrino mass parameter.
It worths noticing that the uncertainty decreases
with increasing true neutrino mass
since a larger mass provides a larger signal and
hence easier to be measured as we already pointed out
above. Comparing with the other constraints (red dashed),
the gravitational focusing measurement is particularly
good for larger mass. Namely, the blue solid lines
shrinks towards the central values (black dotted) much
faster than the red dashed lines. This is because
the cosmic gravitational focusing has fourth power
dependence on the neutrino masses as we emphasized
above while the conventional ways has only linear
dependence. This trend is even more prominent for
the IO case in the right panel with two heavy
neutrinos. It fully demonstrates the advantage of
cosmic gravitational focusing as an extra and
independent cosmological measurement of the neutrino masses.

The synergy (black solid) of cosmic gravitational
focusing and
other cosmological constraints significantly improve
the upper and lower limits across the whole parameter
space. Especially, the intersection of the lower limit
curve with the horizontal axis can move below 30\,meV.
In other words, it is possible to put a lower limit
on the lightest neutrino mass at 95\%\,C.L. for true
values down to 24\,meV (29\,meV) with NO (IO).
Below that, only upper limit is possible which
can be as small as 25\,meV (29\,meV) for NO (IO) when
the lightest mass approaches 0. These features
apply for both NO (left) and IO (right) with slightly
different values. For comparison, the 95\% upper
limit from the combination (grey region) of existing
cosmological observations and neutrino oscillation
experiments is also shown.

\section{Conclusions}
\label{sec:conclusion}

Under the gravitational force of DM overdensity,
the C$\nu$F can experience gravitational lensing
which has two phenomenological consequences of
dynamical friction and gravitational focusing.
While the dynamical friction is mainly the drag
force between C$\nu$F and DM halo, the gravitational
focusing directly leads to a C$\nu$F density enhancement
on the downstream side of the DM halo. This one-sided
density enhancement is essentially a dipole component
after subtracting the average density. Since the C$\nu$F
density can also contribute to the total matter density,
the dipole structure can manifest itself in the galaxy
distribution as an imaginary galaxy cross correlation
in the Fourier space. It is much more intuitive and
appropriate to use cosmic gravitational focusing to
name this effect.

We use both single-particle trajectory with general
relativity for point source of gravitational potential
and Boltzmann equation formalism for continuous overdensities
to describe the cosmic gravitational focusing. Our
results apply for both relativistic and non-relativistic
particles with dependence on the fourth power of mass
and temperature. While the CMB photons have negligible
effect with vanishing mass, the cosmic gravitational
focusing mainly comes from C$\nu$F with at lease one
heavy mass. The cosmic gravitational focusing can then serve
as an independent way of measuring the neutrino masses.

When deriving the galaxy cross correlation, we keep the
most general form of bias for both C$\nu$F and DM
overdensities. Since C$\nu$F and DM have different
clustering properties, it makes much more sense for them
to have different bias. More specifically, the C$\nu$F
bias is very close to 1 with only $\lesssim 2\%$
deviation. This allows two different RSD terms that
add up coherently to enhance the cosmic gravitational
focusing effect in the galaxy cross correlation.
With this enhancement, the cosmic gravitational focusing
provides a more powerful way of determining the neutrino
masses than the mass sum constraint. Since observing
the dipole density distribution requires galaxy
cross correlation among categories with different bias,
we split the BGS galaxies into two sub-categories and
find the optimal mass splitting value to maximize the
SNR.

Finally, we quantitatively show the advantages of
cosmic gravitational focusing by taking the projected
sensitivity at DESI as an illustration. The SNR can
improve by at least a factor of 2 after incorporating
the DESI gravitational focusing measurement to the
existing cosmological constraints from CMB and LSS.
Especially the vanishing lightest neutrino mass with IO
can be excluded up to almost $3 \sigma$ and the upper
limit on the lightest neutrino mass can touch down to
3.2\,meV at 95\%\,C.L. for NO as shown in \gfig{fig:SNR_total}.
Even for the projected sensitivity of CMB and LSS
observations at CSST, DESI-II, EUCLID, and Subaru PFS,
the cosmic gravitational focusing sensitivity is still
very competitive as summarized in \gfig{fig:error_SNR}.
A lower bound is possible for smaller lightest mass than
the other ways. Its advantage is more significant for
heavier neutrinos with fourth power dependence on the
neutrino mass.

\section*{Acknowledgements}

The authors would like to thank Xiao-Hu Yang and Hong-Ming Zhu
for useful discussions.
The authors are supported by the National Natural Science
Foundation of China (12375101, 12090060, 12090064, and 12247141)
and the SJTU Double First Class start-up fund (WF220442604).
SFG is also an affiliate member of Kavli IPMU, University of Tokyo.
PSP is also supported by the Grant-in-Aid for Innovative Areas No.\,19H05810.

\appendix

\section{C$\nu$F-DM Relative Velocity and Its Dispersion}
\label{sec:Vnuc}

The crucial factor of cosmic gravitational focusing, namely
the relative velocity between C$\nu$F and DM halo,
is developed in the evolution of early Universe. In the presence
of matter density fluctuations, the cosmic neutrinos fall into
the gravitational potential well and obtain a relative velocity.
It is then necessary to first estimate the overdensity
fluctuation and the relative velocity therein as we elaborate
in this appendix.

With perturbative overdensity in the early Universe, the linear
approximation \cite{Zhu:2013tma} can apply. As the primordial
fluctuations usually follow gaussian distribution,
the ensemble average of the relative
velocity alone should vanish, $\langle {\bm v}_{\nu c} \rangle = 0$.
In addition, the Universe is homogeneous and isotropic at large
scales which means the relative velocity should appear at small
scales. 
Especially, cosmic neutrinos and DM have a distinct thermal history 
which allows a non-zero relative velocity at scales below
$\sim 100$\,{\rm Mpc} \cite{Zhu:2013tma,Okoli:2016vmd}.
Since 100\,Mpc is already quite sizable, the C$\nu$F-DM
relative velocity distribution can be decomposed into a
background (or bulk) velocity $\bm v^{\rm bg}_{\nu c}$
and small perturbations
$\bm u_{\nu c}(\bm x)\equiv \bm v_{\nu c} - \bm v^{\rm bg}_{\nu  c}$
at even smaller scales. The small perturbations arond the
bulk velocity $\bm v^{\rm bg}_{\nu c}$ are isotropic
$\langle \bm u_{\nu c}(\bm x) \rangle = 0$
and the relative velocity dispersion can be estimated
by the background bulk velocity,
$ \sqrt{\langle \bm v_{\nu c}^2 \rangle}
= \sqrt{\langle (\bm 
v_{\nu c}^{\rm bg})^2 \rangle} + \mathcal O(\bm u_{\nu c}^2)
\approx v_{\nu c}^{\rm bg}$.

We estimate the background velocity as the root-mean-square of 
the relative velocity field \cite{Zhu:2013tma,Okoli:2016vmd},
\begin{align} 
  v_{\nu c}^{\rm bg}
\approx
  \langle \bm v_{\nu c}^2 (R) \rangle  
= 
  \int \frac{d |\bm k|}{|\bm k|} 
  \Delta_\zeta^2 (\bm k)
  \left| \widetilde W (|\bm k| R) \right|^2
  \frac{ | T_{\theta_{\nu}} - T_{\theta_{c}}|^2 }{|\bm k|^2},
\label{eq:v_nuc2}
\end{align} 
where the primordial power 
spectrum $\Delta_\zeta$ and transfer function of the 
velocity divergence  
$T_{\theta_a} \equiv \tilde \theta_a / \zeta$ ($a = \nu,c$)
can be obtained from the CLASS code 
\cite{Blas:2011rf} with the Plank 2018 best fit~\cite{Planck:2018vyg}. 
The top-hat window function $\widetilde W (|\bm k| R)$ 
selects those scales smaller than $R$.

Since the cosmic gravitational focusing is proportional to
the relative velocity, a larger velocity increases the signal.
At the same time, the velocity becomes larger for smaller
filtering scale $R$
\cite{Okoli:2016vmd} and the signal becomes optimal for smaller $R$. 
However, the average distance between two nearest galaxies
is 2\,Mpc \cite{2012arXiv1212.1671W}. To observe the effect
of cosmic gravitational focusing on the dipole galaxy distribution,
it is necessary for the window function to cover neighboring
galaxies with radius $R \gtrsim 2$\,Mpc. In addition,
the nonlinear effect would become important below
$R = |\bm k|^{-1} \sim 5\,{\rm Mpc}/h$ \cite{Dodelson:cosmo2nd}.
In order to use the power spectrum simulated in
the linear regime with typical code such as CLASS\cite{Blas:2011rf},
the window function radius cannot be too small. A reasonable choice is
$R = 5\,{\rm Mpc}/h$.

\begin{figure}[t]
\centering
\includegraphics[width=0.49\textwidth]{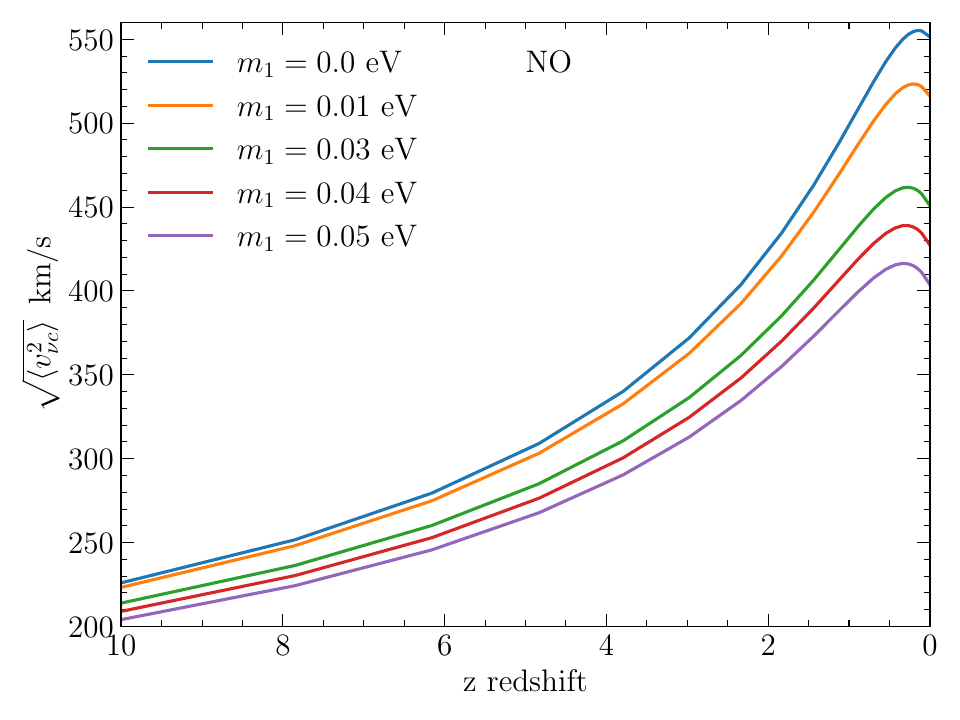}
\includegraphics[width=0.49\textwidth]{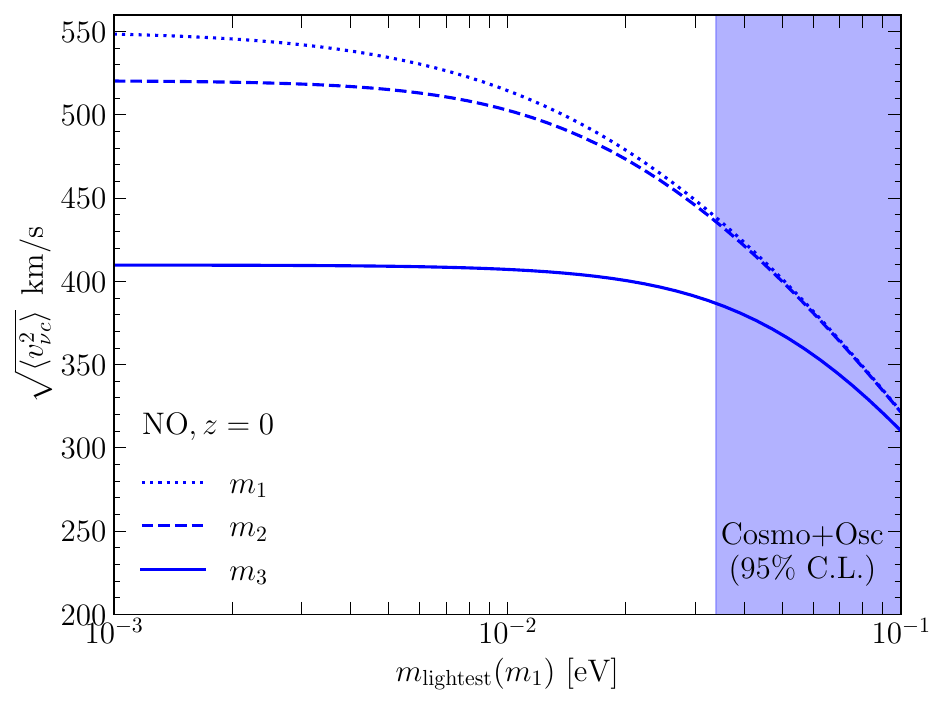}
\caption{(Left) The relative velocity between C$\nu$F and DM
for the lightest mass $m_1$ in the NO as a function
of redshift $z$. (Right) The relative velocities for the three
massive neutrinos as functions of the lightest neutrino mass
($m_{\rm lightest} = m_1$ for NO) at nowadays ($z=0$).
}
\label{fig:Dvvrsk}
\end{figure}

In the left panel of \gfig{fig:Dvvrsk}, 
we show how the relative velocity changes with 
redshift $z$ for different neutrino masses. 
The relative velocity is larger for smaller masses 
and increases with redshift until it reaches 
around $z \sim  0.4$. As explained earlier, the C$\nu$B
neutrinos fall into the gravitational potential of mass
density fluctuations and obtain a relative velocity. Since
the matter clustering increases with the Universe evolution,
or equivalently decreasing redshift, the C$\nu$B-DM
relative velocity also increases and finally reaches
$(400 \sim 500)$\,km/s. After that, the dark energy (DE)
starts to dominate and the relative velocity decreases
since DE can resist the clustering tendency.

For the degenerate case, three neutrinos have the roughly same
relative velocity \cite{Zhu:2013tma}. But for the real
case, the three neutrinos could have different bulk velocities. 
To show this difference, we take the NO for example
in the right panel of \gfig{fig:Dvvrsk}. Since the lightest
neutrino (solid blue) has a larger velocity dispersion of
$v_{\nu c} \propto \sum \int p_\nu / m_\nu$ than that of
the two heavy ones (dashed and dotted blue), it has
a relative velocity about 40\% larger. In this region,
the mass squared differences, $m_2\approx \sqrt{\Delta m_{21}^2}$
and $m_3\approx \sqrt{\Delta m_{31}^2}$, dominate over
the lightest mass $m_{\rm lightest}$. So varying
the tiny $m_{\rm lightest}$ has negligible effect and
the curves are almost flat. As the lightest neutrino mass
increases, the two light neutrino masses $m_1$ and $m_2$
become degenerate first as manifested by the
converging dotted and dashed lines. With further increasing
mass, once the lightest mass surpasses the larger mass squared
difference,
$m_{\rm lightest} = m_1 > \sqrt{\Delta m_{31}^2} \approx 0.05~{\rm eV}$,
all the masses become degenerate, $m_1 \approx m_2 \approx m_3$. 
This explains why the three neutrinos converge on the right-hand
side with roughly the same velocity around $300$\,km/s. 
In all cases, the background neutrino velocities are
nonrelativistic and of order 
$\sqrt{\langle v_{\nu_i c}^2 \rangle}  \sim \mathcal O(100)\,{\rm km/s}$.
Most importantly, the velocity dispersion decreases with the
neutrino mass. This is because the C$\nu$F and DM relative velocity
have been implemented \cite{Ma:1994dv} in the CLASS \cite{Blas:2011rf}, CAMB
\cite{Howlett:2012mh},  CMBEASY \cite{Doran:2003sy},
and CMBFAST \cite{Zaldarriaga:1997va} packages. Although not emphasized,
the cosmic dynamical friction has already been existing in the
early Universe simulations. With larger mass, the drag force
between C$\nu$F and matter overdensity
is also larger to reduce the relative velocity.

\section{Neutrino Dipole Density Distribution Induced by Anisotropic Phase Space Distribution}
\label{app:phase_form}

The C$\nu$F has non-zero velocity relative
to the DM halo. As shown in \gfig{fig:Dvvrsk}, this relative
velocity  
$|\bm v_{\nu c}| \approx 10^{-3}$ (100 km/s) is non-relativistic
even if individual neutrinos can be relativistic.
We can perform Lorentz boost from the C$\nu$F frame
(${\bm p}'$) to the DM frame (${\bm p}$). When expanded to
the linear order of $\bm v_{\nu c}$, the cosmic neutrino
momentum and energy transform as,
\begin{eqnarray}
  \bm p 
\approx 
  \bm p'  + E_{\bm p'} \bm v_{\nu c} 
,~{\rm and} \quad 
  E_{\bm p} 
\approx 
  E_{\bm p'} + \bm p' \cdot \bm v_{\nu c}.
\label{eq:lorentz1st_copy}
\end{eqnarray}
As a result, the isotropic cosmic neutrino phase-space
distribution function in the C$\nu$F frame receives
anisotropic correction to become
$ f_\nu(\bm p_i, \bm v)
\approx
  2 / \left( e^{ | \bm p_i - E_{\bm p_i} \bm v_{\nu c} | / T} + 1 \right)$
in the DM frame as summarized in
\geqn{eq:nuPhase-Space}. The prefactor 2 denotes the
equal contributions of neutrino and anti-neutrino for the
same mass eigenstate. To make the effect of
C$\nu$F-DM relative velocity ${\bm v}_{\nu c}$ explicit,
it is better to expand to the linear order of ${\bm v}_{\nu c}$.
With a nonrelativistic Lorentz boost $|{\bm v}_{\nu c}| \sim 10^{-3}$ as shown in
\gfig{fig:Dvvrsk}, the correction to the matter density
is essentially a Lorentz boost factor
$\gamma \equiv \sqrt{1 - |{\bm v}_{\nu c}|^2}$
which only gives a factor of $10^{-6}$ difference.
So it is safe to work in the DM frame.

Below we show the detailed derivations to fill the gap
between \geqn{eq:drho_nu} and \geqn{eq:drho_nu2}. First,
we need to perform the momentum gradient 
$\nabla_{\bm p}$ on the homogeneous part $\overline f_\nu(\bm p)$
of the neutrino phase space distribution function,
\begin{eqnarray} 
  \frac{\partial \overline f_\nu(\bm p) }{\partial p_j} 
=
  \frac{\partial p'_i }{\partial p_j} 
  \frac{\partial \overline f_\nu(|\bm p'|) }{\partial p'_i}
=
  \frac{\partial p'_i }{\partial p_j} 
  \hat p'_i
  \frac{d \overline f_\nu(|\bm p'|) }{d |\bm p'|}.
\end{eqnarray} 
Using the Lorentz transformation in \geqn{eq:lorentz1st_copy}, 
the first derivative can be written in terms of the neutrino
momentum and energy,
$ \partial p'_i/ \partial p_j 
\approx \delta_{ij} - p_j v_{\nu c i}/E_{\bm p}
\approx \delta_{ij} - p'_j v_{\nu c i}/E_{\bm p'} $.
Putting things together, the momentum gradient becomes,
\begin{align} 
  \nabla_{\bm p} \overline f_\nu (\bm p) 
\approx
  \hat {\bm p'}
  \left( 
    1
  -
  \frac{ \bm v_{\nu c} \cdot \bm p' }{E_{\bm p'}}
  \right)
  \frac{d \overline f_\nu(|{\bm p'}|) }{d |{\bm p'}|}.
\label{eq:nablafpp}
\end{align} 

To implement the Lorentz transformation, one may first
use \geqn{eq:lorentz1st_copy} and \geqn{eq:nablafpp}
to replace the neutrino momentum $\bm p$ (also $E_{\bm p}$)
and its gradient in \geqn{eq:drho_nu},
\begin{align}
  \delta \widetilde \rho_\nu
\approx &
  \widetilde \Psi 
  \int \frac{d^3 \bm p' }{(2 \pi)^3}
   E_{\bm p'}
\left(
    1
  + \frac{ \bm p' \cdot \bm v_{\nu c} }{E_{\bm p'}}
\right)
\\ 
&  \times 
\left\{
  \frac{
      m^2_\nu 
   +  2  |\bm p'|^2 
   + 4 E_{\bm p'} ( \bm p' \cdot \bm v_{\nu c} ) }
  { ( \bm p' + E_{\bm p'} \bm v_{\nu c} ) \cdot \bm k}
  \bm k \cdot  \hat{\bm p'} 
  - 
  \left[|\bm p'| + E_{\bm p'} (\bm v_{\nu c} \cdot \hat{\bm p'}) \right]
\right\}
  \frac{d \overline f_\nu (|\bm p'|)}{d |\bm p'|},
\notag
\end{align} 
where the integration variable has also been replaced
accordingly,
$d^3 \bm p/ E_{\bm p}  = d^3 \bm p'/E_{\bm p'}$.
It is much more convenient to see the features by 
combining the two terms in the curly bracket and
extracting the denominator as an overall factor,
\begin{align} 
\hspace{-1mm}
  \delta \widetilde \rho_\nu
\approx
  \widetilde \Psi
  \int \frac{d^3 \bm p' }{(2 \pi)^3}
  \frac{d \overline f_\nu}{d |\bm p'|} 
  \frac{
   E_{\bm p'}^3}{( \bm p' + E_{\bm p'} \bm v_{\nu c} ) \cdot \bm k}
\left\{
    \bm k \cdot \hat {\bm p'} 
+
  \frac{|\bm p'|}{E_{\bm p'}}
  \left[  
    4 ( \hat{\bm p'} \cdot \bm v_{\nu c})
    (\bm k \cdot \hat{\bm p'}) 
    -
    (\bm v_{\nu c} \cdot \bm k)
  \right]
\right\}.
\label{eq:RelrhoInt_app}
\end{align}
Of the three terms in the curly bracket, the first one
dominates for non-relativistic neutrinos
$E_{\bm p'} \gg |\bm p'|$ while
the second term is suppressed by $|\bm p'|/E_{\bm p'}$.

Since the relative bulk velocity ${\bm v}_{\nu c}$ is
non-relativistic, the denominator can vanish and lead
to a divergence which can be replaced by a $\delta$-function
using the SPW theorem
\cite{Weinberg:1995mt}, 
\begin{eqnarray}
    \frac 1 {( \bm p' + E_{\bm p'} \bm v_{\nu c} ) \cdot \bm k}
\quad \Rightarrow \quad 
  \mathcal P
  \left(
    \frac 1 {( \bm p' + E_{\bm p'} \bm v_{\nu c} ) \cdot \bm k} \right) 
+
  i \pi 
  \delta(\bm p'\cdot \bm k + E_{\bm p'} \bm v_{\nu c} \cdot \bm k).
\end{eqnarray}
For given bulk velocity ${\bm v}_{\nu c}$ and comoving
wave number $\bm k$, the argument of the $\delta$-function
depends on the openning angle between ${\bm p}'$ and $\bm k$.
It is then more convenient to rewrite the integration
$d^3 {\bm p}' = |{\bm p}'|^2 d |{\bm p}'| d \Omega_{\bm p'}$
in terms of its solid angle. Then the $\delta$-function
can be rewritten in terms of openning angles,
$\delta(|{\bm p'}||{\bm k}| \cos \theta + E_{\bm p'} {\bm v}_{\nu c} \cdot {\bm k}) 
=
  \delta(\cos \theta - \cos \theta_0) / |\bm p'||{\bm k}|
$
with
$\cos \theta_0 \equiv - ({\bm v}_{\nu c} \cdot \hat {\bm k}) / |{\bm v}'|$.
Note that ${\bm v}' \equiv |{\bm p}'| / E_{\bm p'}$
is the velocity of individual neutrino in the C$\nu$F frame.
The integration over $\delta$-function is non-zero only if 
$|\bm p'|  > E_{\bm p'} |{\bm v}_{\nu c} \cdot \hat {\bm k}|$.
It is interesting to see this gives an imaginary part to
the neutrino density distribution in the momentum space,

The ${\bm k} \cdot \hat{\bm p}'$ in the first two terms of
\geqn{eq:RelrhoInt_app} has been replaced by
$- ({\bm v}_{\nu c} \cdot {\bm k}) / |{\bm v}'|$
due to the $\delta$-function.
\begin{align} 
  {\rm Im}[\delta \rho_\nu]
= &
- \frac{\widetilde \Psi}{8 \pi^2}
  \int d|\bm p'| 
  |\bm p'|^2 E_{\bm p'}^3
  \frac{d \overline f}{d |\bm p'|}
  \Theta(|\bm p'| - E_{\bm p'}|{\bm v}_{\nu c} \cdot \hat {\bm k}|)
\nonumber
\\ \times &
  \frac{2 \pi}{|\bm p'| |\bm k| } 
\left[
    \frac{E_{{\bm p}'}}{|{\bm p}'|} 
    ({\bm v}_{\nu c} \cdot {\bm k})
+ 4  \frac{E_{\bm p'}}{ |\bm p'|}
  ( \bm v_{\nu c} \cdot \hat{\bm k} )^3 |\bm k|
+
  \frac{|{\bm p}'|}{E_{{\bm p}'}}
  ({\bm v}_{\nu c} \cdot {\bm k})
\right]. 
\end{align}
Although the first two terms share the same
$E_{\bm p'}/{\bm p'}$ factor, the second one is suppressed
by $v^3_{\nu c}$ and hence can be omitted for simplicity.
The remaining two terms can be
comparable for relativistic neutrinos. The inner product
${\bm v}_{\nu c} \cdot {\bm k}$ can be extracted as an
overall factor to give a much simpler form,
\begin{align} 
  {\rm Im}[\delta \rho_\nu]
=
-  \frac{\widetilde \Psi ({\bm v}_{\nu c} \cdot \hat {\bm k})}{4\pi}
  \int d|\bm p'| 
  \left(
    m_\nu^4
  +3m_\nu^2|\bm p'|^2
  +2 |\bm p'|^4
    \right)
  \frac{d \overline f}{d |\bm p'|}
    \Theta(|\bm p'| - E_{\bm p'} |{\bm v}_{\nu c} \cdot \hat {\bm k}|).
\label{eq:Imrho}
\end{align}
Similar to the classical picture with a point source
of gravitational potential, the cosmic gravitational focusing
also relies on the fourth power of neutrino mass $m_\nu$
and its momentum $\bm p'$. The non-relativistic neutrino
case is dominated by the mass $m^4_\nu$ while the relativistic
case by the neutrino momentum.

The momentum integration will give the dependence on the
neutrino temperature $T_\nu$. Note that the integration
range is determined by the $\Theta (|\bm p'| - E_{\bm p'} |{\bm v}_{\nu c} \cdot \hat {\bm k}|)$
which means $
|\bm p'| 
> m_\nu |\bm v_{\nu c} \cdot \hat{\bm k}|
  /\sqrt{1 - (\bm v_{\nu c} \cdot \hat{\bm k})^2} 
\approx m_\nu |\bm v_{\nu c} \cdot \hat{\bm k}| $.
The integration in \geqn{eq:phiGeneral} can be generally
parametrized as, 
\begin{align}
  \int d|\bm p'| 
  |\bm p'|^{2n}
  \frac{d \overline f}{d |\bm p'|}
    \Theta(|\bm p'| - E_{\bm p'}|{\bm v}_{\nu c} \cdot \hat {\bm k}|)
=
  2 T_\nu^{2n}  
  \int_{y_0}^\infty dy 
  y^{2n}
  \frac{d [e^{y} + 1]^{-1}}{\partial y}
\equiv 
  T_\nu^{2n}  f_n(y_0),
\label{eq:dfdpTheta}
\end{align}
by performing a 
variable change $y \equiv |
{\bm p'}|/T_\nu$  with $y_0 \equiv $ 
$m_\nu |{\bm v}_{\nu c} \cdot \hat {\bm k}|/T_\nu$. 
The factor 2 comes from the phase space distribution function
to account for both neutrino and antineutrino contributions.
For convenience, we have defined three handy functions
$f_n(y_0)$ $(n=1,2,3)$, 
\begin{align}
  f_n (y_0)
\equiv
  2 \int_{y_0}^\infty dy 
  y^{2n}
  \frac{d [e^{y} + 1]^{-1}}{\partial y}
= 
- \int_{y_0}^\infty dy 
  y^{2n}
  \frac 1 {1 + \cosh y}.
\end{align}
Finally, \geqn{eq:Imrho} can be written as,
\begin{align} 
  {\rm Im}[\delta \rho_\nu]
=
-  \frac{\widetilde \Psi }{4\pi}
   \sum_i 
   ({\bm v}_{\nu_i c} \cdot \hat{{\bm k}})
   \left[
     m_i^4 f_0 
   + 3 m_i^2 T^2_\nu f_1 
   + 2 T_\nu^4 f_2  
   \right],
\label{eq:deltarho_t}
\end{align}
where we have summed over the three mass eigenstates.
The same precedure is also applied from \geqn{eq:phiGeneral} to \geqn{eq:phiRel}.
The equal contributions of neutrino and antineutrino
for the same mass eigenstate has been taken into
account by the factor of 2 intrinsically defined
in the neutrino phase space distribution functions
at the beginning of this appendix.

The largest $y_0 \approx 0.5$ happens for
$m_3 \lesssim 0.07$\,eV (NO for $\sum m_i = 0.13\,{\rm eV}$) 
when $T_\nu\sim 1.7\times 10^{-4}$\,eV. 
The function $f_n (y_0)$ can be expanded as power
series of $y_0$,
\begin{align}
  f_0 (y_0) 
\approx
  - 1 + \frac{y_0} 2 + \mathcal O(y_0^3) ,
\quad
  f_1 (y_0)
\approx
  -\frac{\pi^2} 3 + \mathcal O(y_0^3) ,
~{\rm and} \quad
  f_2 (y_0)
\approx
  -\frac{7 \pi^4}{15} + \mathcal O(y_0^5).
\label{eq:f1f2f3_app}
\end{align}
For $f_0$, we preserve the linear order of $\mathcal O(y_0)$, 
since the difference between $-1 + y_0/2$ and $-1$ 
is almost $25\%$ for the largest $y_0 \approx 0.5$.
The next order $\mathcal O(y_0^3)$ correction 
is smaller than $0.7\%$ and can be safely ignored.
With such an approximation, 
\geqn{eq:deltarho_t} becomes,
\begin{align}
  {\rm Im}[\delta \rho_\nu]
=
  \frac{\widetilde \Psi }{4\pi}
  \sum_i 
   ({\bm v}_{\nu_i c} \cdot \hat{{\bm k}})
\left[
    m_i^4 
    \left(1 - \frac{m_i |{\bm v}_{\nu c} \cdot \hat {\bm k}|}{2T_\nu}
  \right)
+ \pi^2 m_i^2 T_\nu^2  
+\frac{14}{15} \pi^4 T_\nu^4 
\right].
\label{eq:f1f2f3}
\end{align}
For the second term in the parenthesis, the non-relativistic
C$\nu$F-DM relative velocity is amplified by the prefactor,
$m_1 / 2 T_\nu \approx \mathcal O(100)$. The gravitational
potential $\widetilde \Psi$ is generated mainly by matter and
hence should be replaced by the matter density, $\widetilde \Psi 
= - 4 \pi G a^2 \rho_m \widetilde \delta_{m0}/|{\bm k}|^2$,
according to the gravitational Poisson equation.

\section{Imaginary Density Fluctuation as Ensemble Average of Primordial Power Spectrum}
\label{app:average}

The neutrino energy density also contributes to the total
matter overdensity 
$ \widetilde \delta_m
\equiv 
  \widetilde \delta_{m 0}
  (1 + i \widetilde \phi)$,
as mentioned in \gsec{sec:deltam}.  
Starting from \geqn{eq:TildephiRel} whose derivation details
have been provided in the previous \gapp{app:phase_form}, the imaginary
overdensity can be parametrized as a phase shift to the
matter overdensity,
\begin{align}
  \widetilde \phi_i
\equiv
-  \frac{a^2}{|\bm k|^2} 
  (\bm v_{\nu_i c} \cdot \hat{\bm k})
\left(
  \beta_i 
- \alpha_i |\bm v_{\nu c} \cdot \hat {\bm k}|
\right),
\label{eq:tildephi_aux}
\end{align}
where we define two coefficients $\beta_i$ and $\alpha_i$,
\begin{align}
  \beta_i 
\equiv 
  G
\left(
  m_i^4 
+ \pi^2 m_i^2 T_\nu^2  
+\frac{14}{15} \pi^4 T_\nu^4 
\right),
\quad 
  \alpha_i
\equiv 
  G \frac{m_i^5}{2 T_{\nu}},
\end{align}
for simplicity. While the $\beta_i$ term associated
with a linear $\bm v_{\nu_i c} \cdot \hat{\bm k}$
captures the main contributions from the neutrino mass
eigenstate with mass $m_i$, the $\alpha_i$ term has
quadratic dependence on $\mathcal O(\bm v_{\nu c}^2)$.
In other words, the $\alpha_i$ term originates from
the $m_i |{\bm v}_{\nu c} \cdot \hat {\bm k}| / 2 T_\nu$
term of \geqn{eq:f1f2f3}. Since the largest mass is at least
$m_1 (m_3) \gtrsim 0.05$\,eV while the neutrino
temperature nowadays is around $10^{-4}$\,eV,
$\alpha_i / \beta_i \approx m_i / 2 T_\nu
\gtrsim 250$ and $\alpha_i v_{\nu c} / \beta_i \gtrsim 0.25$.
The $\alpha_i$ term can contribute at least 25\%
for the heaviest neutrino and it is necessary to
expand to the second order of the $\alpha_i$ terms.

Putting \geqn{eq:tildephi_aux} back, the two ensemble
averages 
$\langle \widetilde{\phi}^2_i \rangle$ and $\langle \dot{\widetilde \phi^2_i} \rangle$ in
\geqn{eq:lnl_dpdppp} are,
\begin{subequations}
\begin{align}
\hspace{-4mm}
  \langle \widetilde{\phi}^2_i \rangle 
= &
  \frac{a^4}{|\bm k|^4} 
  \langle (\bm v_{\nu_i c} \cdot \hat{\bm k})^2\rangle 
\left[ 
   \beta_i^2 
  - \frac{4\sqrt 2}{\sqrt\pi} 
    \alpha_i \beta_i
    \sqrt{\langle (\bm v_{\nu_j c} \cdot \hat{\bm k})^2\rangle }
  + 3 \alpha_i^2
    \langle (\bm v_{\nu_j c} \cdot \hat{\bm k})^2\rangle 
\right],
\label{eq:phiphi}
\\
\hspace{-4mm}
  \frac{\langle \dot{\widetilde \phi^2_i} \rangle}{H^2} 
= &
  \frac{a^2 \beta_i^2}{|\bm k|^4}
\left[
  4 a^2
\left\langle 
  (\bm v_{\nu_i c} \cdot \hat{\bm k})^2
\right\rangle   
  - 4 a 
\left\langle 
  (\bm v_{\nu_i c} \cdot \hat{\bm k})
  \partial_z (\bm v_{\nu_i c} \cdot \hat{\bm k})
\right\rangle
  +
\left\langle 
  [\partial_z (\bm v_{\nu_i c} \cdot \hat{\bm k})]^2
\right\rangle   
\right]
- 
  \frac{ 4 \sqrt 2 a^2 \beta_i \alpha_i}{\sqrt \pi |\bm k|^4}
\notag
\\ 
&  \times
\left[
       6 a^2 
\left\langle 
      (\bm v_{\nu_i c} \cdot \hat{\bm k})^2
\right\rangle   
    -  7 a 
\left\langle 
      (\bm v_{\nu_i c} \cdot \hat{\bm k})
      \partial_z (\bm v_{\nu_i c} \cdot \hat{\bm k})
\right\rangle   
    +  2
\left\langle 
      [\partial_z (\bm v_{\nu_i c} \cdot \hat{\bm k})]^2
\right\rangle   
\right]
  \sqrt{\langle  (\bm v_{\nu_i c} \cdot \hat{\bm k})^2 \rangle}
\notag
\\
& +
  \frac{a^2 \alpha_{\nu_j}^2}{|\bm k|^4} 
\left\{ 
  27 a^2 
\left\langle 
  (\bm v_{\nu_j c} \cdot \hat{\bm k})^2
\right\rangle^2
- 36 a
\left\langle 
  (\bm v_{\nu_j c}\cdot \hat{\bm k})^2
\right\rangle
\left\langle 
  (\bm v_{\nu_j c}\cdot \hat{\bm k}) \partial_z (\bm v_{\nu_j c} \cdot \hat{\bm k})
\right\rangle
\right.
\notag
\\
& + \left.
   4
\left\langle 
  (\bm v_{\nu_j c}\cdot \hat{\bm k})^2 
\right\rangle   
\left\langle 
  [\partial_z (\bm v_{\nu_j c}\cdot \hat{\bm k})]^2
\right\rangle   
+8
\left\langle 
  (\bm v_{\nu_j c}\cdot \hat{\bm k}) 
  [\partial_z (\bm v_{\nu_j c}\cdot \hat{\bm k})]
\right\rangle^2
\right\},
\label{eq:phi2_H2}
\end{align} 
\end{subequations}
where $\partial_z$ is the redshift derivative.
For the heavy neutrino case with
$m_i \gg T_\nu$, the $\alpha_i$
and $\beta_i$ parameters have hierarchical structure,
$\alpha_i = \dot \alpha_i / H \gg \beta_i \gg |\dot \beta_i|/ H$ with
$\dot \beta_i/H = - G \left[ 2 \pi^2 m^2_i T^2_\nu + (56/15) \pi^4 T^4_\nu \right]$.
Comparing the $\beta_i$ and $\dot \beta_i$
terms, $|\dot \beta_i| / \beta_i H
\approx 2 \pi^2 T^2_\nu / m^2_i \lesssim 8 \times 10^{-5}$ which means it is safe to omit the
$\dot \beta_i$ terms. For convenience, the $\dot \alpha_i$
terms in \geqn{eq:phi2_H2} have been replaced by
$\alpha_i H$.

The ensemble averages can now be replaced by the
velocity ensemble averages,
\begin{subequations}
\begin{align} 
\langle 
  (\bm v_{\nu c}\cdot \hat{\bm k})^2 
\rangle
= &
\hspace{1mm}
  \frac 1 3
  \int \frac{d |\bm k'|}{|\bm k'|}
  \Theta (|\bm k| - |\bm k'|) 
  \left| \widetilde W (|\bm k'| R) \right|^2
  \Delta_{\zeta}^2 (\bm k')
  \left| \frac{T_{\theta_{\nu_i c}} (\bm k', z)}{|\bm k'|} \right|^2,
\label{eq:phi2_theta}
\\
\langle 
  (\bm v_{\nu_i c}\cdot \hat{\bm k}) 
  [\partial_z (\bm v_{\nu_i c}\cdot \hat{\bm k})]
\rangle
= & 
\hspace{1mm}
  \frac 1 3
  \int \frac{d |\bm k'|}{|\bm k'|}
  \Theta (|\bm k| - |\bm k'|) 
  \left| \widetilde W (|\bm k'| R) \right|^2
  \Delta_{\zeta}^2 (\bm k')
\left[ 
  \frac{T_{\theta_{\nu_i c}}}{|\bm k'|}
  \frac{\partial_z T_{\theta_{\nu_i c}}}{|\bm k'|}
\right],
\label{eq:vpzv}
\\
\langle 
  [\partial_z (\bm v_{\nu_i c}\cdot \hat{\bm k})]^2
\rangle
= &
\hspace{1mm}
  \frac 1 3
  \int \frac{d |\bm k'|}{|\bm k'|}
  \Theta (|\bm k| - |\bm k'|) 
  \left| \widetilde W (|\bm k'| R) \right|^2
  \Delta_{\zeta}^2 (\bm k')
\left[ 
  \frac{\partial_z T_{\theta_{\nu_i c}}}{|\bm k'|}
\right]^2,
\label{eq:pzv2}
\end{align}
\label{eq:vnuc_Theta}
\end{subequations}
where the relative transfer function 
$ T_{\theta_{\nu_i c}} \equiv T_{\theta_{\nu_i}} - T_{\theta_c}$
is defined as the difference between the C$\nu$F
and DM ones. The Heaviside-$\Theta$ function appears here
to extract only the large scale modes ($|\bm k| \rightarrow 0$) 
that can contribute coherently while those small scale
modes are averaged out \cite{Okoli:2016vmd}. 
As shown in \gfig{fig:SNR_kmode}, the largest SNR happens in
the region of $ k < 0.1\,h/{\rm Mpc}$ where the C$\nu$B-DM power
spectrum predicted by the linear theory receives little suppression 
compared to the $N$-body simulation \cite{Inman:2015pfa}.
Hence, our estimation based on the linear theory is also safe.

\end{document}